\documentclass[12pt]{article}
\usepackage{amssymb}
\usepackage{amsbsy}
\usepackage{verbatim}
\usepackage{youngtab}
\makeatletter \@addtoreset{figure}{section}
\def\thefigure{\thesection.\@arabic\c@figure}
\def\fps@figure{h, t}
\@addtoreset{table}{bsection}
\def\thetable{\thesection.\@arabic\c@table}
\def\fps@table{h, t}
\@addtoreset{equation}{section}

\newtheorem{corollary}{Corollary}[section]
\newtheorem{definition}{Definition}[section]
\newtheorem{theorem}{Theorem}[section]
\newtheorem{proposition}{Proposition}[section]

\newtheorem{examps}{Examples}[section]

\newtheorem{lemma}{Lemma}[section]
\newtheorem{remark}{Remark}[section]
\newtheorem{remarks}[remark]{Remarks}

\def\bx{\begin{example}}
\def\ex{\end{example}}
\def\bxs{\begin{examps}. \rm\begin{enumerate}}
\def\exs{\end{enumerate}\end{examps}}
\def\bd{\begin{definition}}
\def\ed{\end{definition}}
\def\bt{\begin{theorem}}
\def\et{\end{theorem}}
\def\bp{\begin{proposition}}
\def\ep{\end{proposition}}
\def\bc{\begin{corollary}}
\def\ec{\end{corollary}}
\def\bl{\begin{lemma}\em}
\def\el{\end{lemma}}
\def\be{\begin{equation}}
\def\ee{\end{equation}}
\def\bproof{\begin{proof}}
\def\eproof{\end{proof}}
\def\br{\begin{remark}\rm\small}
\def\er{\end{remark}}
\def\brs{\begin{remarks}.\\ \rm\
\begin{enumerate}}
\def\ers{\end{enumerate}\end{remarks}}
\def\bea{\begin{eqnarray}}
\def\eea{\end{eqnarray}}

\def\ra{{\rightarrow}}

\def\mt{{\mapsto}}

\def\det{\mathrm {det}}
\def\Det{\mathrm {Det}}

\def\span{\mathrm {span}}
\def\Fr{\mathrm {Fr}}
\def\Gr{\mathrm {Gr}}
\def\res{\mathop{\mathrm {Res}}\limits}

\def\&{&{\hskip -20pt}}

\def\AA{\mathcal{A}}

\def\DD {\mathcal{D}}
\def\FF{\mathcal{F}}

\def\HH{\mathcal{H}}
\def\JJ{\mathcal{J}}

\def\LL{\mathcal{L}}

\def\OO{\mathcal{O}}

\def\SS{\mathcal{S}}

\def\Cb{\mathbf{C}}
\def\Ib{\mathbf{I}}
\def\Jb{\mathbf{J}}

\def\Tb{\mathbf{T}}
\def\eb{\mathbf{e}}
\def\fb{\mathbf{f}}

\def\hb{\mathbf{h}}
\def\kb{\mathbf{k}}

\def\rb{\mathbf{r}}
\def\tb{\mathbf{t}}
\def\vb{\mathbf{v}}
\def\ub{\mathbf{u}}
\def\Nb{\mathbf{N}}
\def\Pb{\mathbf{P}}
\def\Rb{\mathbf{R}}
\def\Ub{\mathbf{U}}

\def\Zb{\mathbf{Z}}

\def\Hb{\mathbf{H}}

\def\0b{\mathbf{0}}

\def\Cbb{\mathbb{C}}

\def\grA{\mathfrak{A}} \def\gra{\mathfrak{a}}
\def\grB{\mathfrak{B}} \def\grb{\mathfrak{b}}

\def\grP{\mathfrak{P}}

\def\grS{\mathfrak{S}} 
\def\grT{\mathfrak{T}}

\def\grGl{\mathfrak{Gl}}

\def\nchi{\hbox{\raise 2.5pt\hbox{$\chi$}}}

\date{}
\begin{document}
\baselineskip 16pt
\begin{flushright}
CRM-3309 (2010)
\end{flushright}
\medskip
\begin{center}
\begin{Large}\fontfamily{cmss}
\fontsize{17pt}{27pt}
\selectfont
\textbf{Schur function expansions of KP tau functions associated to algebraic curves}\footnote{Work supported in part by the Natural Sciences and Engineering Research Council of Canada (NSERC) and the Fonds FCAR du Qu\'ebec.}
\end{Large}\\
\bigskip
\begin{large}  { {V. Enolski}$^{\star}$\footnote{vze@ma.hw.ac.uk}
 and J. Harnad}$^{\dagger \ddagger}$\footnote{harnad@crm.umontreal.ca}
\end{large}
\\
\bigskip
\begin{small}
$^{\dagger}$ {\em Centre de recherches math\'ematiques,
Universit\'e de Montr\'eal\\ C.~P.~6128, succ. centre ville, Montr\'eal,
Qu\'ebec, Canada H3C 3J7} \\
\smallskip
$^{\ddagger}$ {\em Department of Mathematics and
Statistics, Concordia University\\
7141 Sherbrooke W., Montr\'eal, Qu\'ebec,
Canada H4B 1R6} \\
\smallskip
$^{\star}$ {\em Institute of Magnetism NASU\\
Vernadsky Blvd. 36-b,
Kyiv 03142, Ukraine } \\
\end{small}
\end{center}
\bigskip
\bigskip
\begin{center}{\bf Abstract}
\end{center}
\smallskip

\begin{small}
The Schur function expansion of  Sato-Segal-Wilson KP $\tau$-functions is reviewed. The case of
 $\tau$-functions related to algebraic curves  of arbitrary genus is studied in detail. Explicit
 expressions for the Pl\"ucker coordinate coefficients appearing in the expansion are obtained in terms of  directional derivatives of the Riemann theta function or Klein sigma function along the KP flow directions.  Using the fundamental bi-differential,  it is shown how the coefficients can be expressed as polynomials in terms of Klein's higher genus generalizations of Weierstrass' $\zeta$ and $\wp$ functions. The cases of genus two hyperelliptic and genus three trigonal curves are detailed as illustrations of the approach developed here.
\bigskip \bigskip \eject
\tableofcontents
\end{small}
\bigskip


\section{Introduction}
In the mid-1970s Novikov and Dubrovin (\cite{nov74}, \cite{dn74}, \cite{dub75}),  Its and Matveev (\cite{im75}, \cite{im76}) and others (\cite{ma74}, \cite{la75}, \cite{mvm75}, \cite{dt76}) applied the Lax pair (isospectral deformation) approach to Hill's operator for periodic potentials with finite band spectrum and thereby determined finite gap periodic solutions to the KdV equation  
\be
\mathcal{U}_t=6\,\mathcal{U}\mathcal{U}_x-\mathcal{U}_{xxx}.\label{KdV}
\ee
Its and Matveev  discovered a remarkable formula that provides periodic and, more generally, quasi-periodic solutions of the KdV equation explicitly as a second logarithmic derivative of the Riemann theta-function:
\be
\mathcal{U}(x,t)=-\frac{\partial^2}{\partial x^2 }  \,\mathrm{ln} \,
\theta(\mathbf{U}x+\mathbf{V}t+ \mathbf{W})+C \label{kdvsol}
\ee
with $\Ub,\mathbf{V},\mathbf{W}=\mathrm{const}\in\mathbb{C}^g$ and $C\in \mathbb{C}$. The theta-function appearing here is determined by the period lattice of a hyperelliptic curve $X$ of arbitrary genus $g$, and the ``winding vectors'' $\mathbf{U}, \mathbf{V}$ are periods of abelian differentials of the second
kind. This formed an important part of the general theory of algebro-geometric solutions of the KdV equation (see e.g.  (\cite{dmn76} for a review).  Krichever \cite{Kr77} extended these considerations more generally to  quasi-periodic solutions of the KP hierarchy. This led to a general method of integration of such  partial differential equations determined by the specification of an algebraic curve, and certain additional data on it.  (See(\cite{dub81}) for an overview of this approach and further applications.) These results had a great influence on subsequent developments in the theory of integrable nonlinear hierarchies, and their applications in various domains of mathematics and physics.

The phenomenon of algebro-geometric integrability has been considered from different viewpoints. In this paper we discuss the theory of tau-functions as initiated by Sato  \cite{sato80,sato81,sato82} and  developed in the works Sato, Date, Jimbo, Kashiwara, Miwa and others (see e.g.~\cite{djkm83}). We also make use of the geometrical formulation of Segal and Wilson \cite{sw85}. In this approach the tau-function
 \be
 \tau=\tau_w(\tb), \quad \tau_w(\mathbf{0})\neq 0
 \ee
is understood to depend on two sets of variables, an infinite dimensional vector $\tb=(t_1,t_2,\ldots)\in \mathbb{C}^{\infty}$ and an element $w\in Gr_{\HH_+}(\HH)$ of an infinite dimensional Grassmannian consisting of subspaces of a polarized Hilbert space (the direct sum of two mutually orthogonal subspaces of essentially equal size)
\be
\HH= \HH_+ + \HH_-
\ee
that are ``commensurable'' with the fixed subspace $\HH_+$ (i.e., orthogonal projection to
$\HH_+$ is ``large'' - a Fredholm operator - while projection to $\HH_-$ is,  e.g., of Hilbert-Schmidt class). 
The Pl\"ucker relations, defining the embedding of  $Gr_{\HH_+}(\HH)$ as a subvariety of the  projectivized infinite exterior space $\Pb( \Lambda \HH)$ are equivalent to an infinite set of bilinear differential relations in the ${\tb }$ variables, the Hirota equations which, in turn imply the equations of the KP hierarchy. 

In the special case to be considered here, $w$ is determined by certain algebro-geometric data, the Krichever-Novikov-Dubrovin (KND) data, consisting of an algebraic curve $X$ of genus $g$, and a non-special positive divisor of degree $g$:
\be
\DD = \sum_{i=1}^g p_i, \qquad p_i\in X,
\ee
a point $p_\infty$ identified as ``infinity'' and a local parameter $\xi = \frac{1}{z}$ with $\xi(p_\infty)=0$.

We further make a choice of a homology basis  $\{\mathfrak{a}_1, \dots, \mathfrak{a}_g; \mathfrak{b}_1, \dots , \mathfrak{b}_g\}$ for $X$ consisting of $\mathfrak{a}$ and $\mathfrak{b}$ cycles satisfying the intersection relations
\be
\mathfrak{a}_i \circ \mathfrak{a}_j =0, \quad \mathfrak{b}_i \circ \mathfrak{b}_j = 0, \quad \mathfrak{a}_i \circ \mathfrak{b}_j = \delta_{ij},
\ee
a normalized basis $\{u_1, \dots , u_g\}$ of holomorphic abelian differentials,
and a canonical polygonization of $X$ obtained by cutting along the $\mathfrak{a}$ and $\mathfrak{b}$-cycles. Selecting an arbitrary base point  $p_0$   the Abel map,
 \be
\AA: (\SS^g X) \ra \JJ(X) = \Cbb^g/\Gamma
\ee
from the $g$th symmetric power of $X$ to its Jacobian variety (the quotient of $\Cbb^g$ by the lattice of periods) is defined by integration of these differentials from the base point to the points in question. The KND data determine a vector $\eb=(e_1,\ldots,e_g)^T \in\mathbb{C}^g$ as the image  of the divisor $\DD$  under the Abel map (within the polygonization)
\be
\eb =\AA(\DD) -  \AA(p_\infty) + \mathbf{K}
\ee
translated by the Riemann constant $\mathbf{K}$ corresponding to  the polygonization.  We may then denote, in short, the corresponding $\tau$-function as $\tau(\eb, \tb)$.

The tau function can be represented as a Taylor series in $\tb$ and then re-expressed as an expansion in a basis consisting of  Schur function $s_{\lambda}(\tb)$, labelled by partitions,  $\lambda=(\lambda_1,\lambda_2,\ldots \lambda_{ \ell(\lambda)})$, (where the $\lambda_i$'s form a weakly decreasing sequence of non-negative integers $\lambda_1 \ge \lambda_2 \ge \dots$, with the last nonzero term $\lambda_{\ell(n)}$,  where  $\ell (\lambda)$ is the {\it length} of $\lambda$). The flow parameters $(t_1, t_2, \dots )$ may be identified with monomial sums
\be
t_i = {1\over i} \sum_{a=1}^N x_a^i
\ee
in terms of $N$ auxiliary variables for any $N\ge \ell(\lambda)$, by taking the (stable) limit $N\ra \infty$.

The Cauchy-Littlewood identity (\cite{mac90}) (or equivalently, the abelian group property of the KP flow flows) permits us to express this expansion in the form
\be
\tau(\eb,\tb)=\sum_{\lambda}\left.\left[ s_{\lambda}\left( \frac{\partial}{\partial t_1},
\ldots, \frac{1}{k} \frac{\partial}{\partial t_k},\ldots  \right) \tau(\eb,\tb)\right]\right|_{\tb=\mathbf{0}} s_{\lambda}(\tb),   \label{tauexpan}
\ee
where the summation runs over all partitions $\lambda$. The important point is that the coefficients in this
expansion
\be
\pi_\lambda(w) :=  s_{\lambda}\left( \frac{\partial}{\partial t_1},
\ldots, \frac{1}{k} \frac{\partial}{\partial t_k},\ldots  \right) \tau(\eb,\tb)|_{\tb=\mathbf{0}}
\label{plucker_tauderiv}
\ee
are just the Pl\"ucker coordinates of the element $w\in \Gr_{\HH_+}(\HH)$ under the Pl\"ucker embedding
\be
\grP : \Gr_{\HH_+}(\HH)\ra \Pb(\FF)
 \ee
 into the projectivization of the exterior space
 \be
 \FF= \Lambda \HH ,
 \ee
which is a completion of the space of sums over a basis consisting of semi-infinite wedge products of basis elements of $\HH$ (the Fermionic Fock space). In this setting, the Hirota bilinear relations of the KP hierarchy are simply equivalent to the Pl\"ucker relations satisfied by the coefficients $\{\pi_\lambda(w)\}$.

To each partition $\lambda=(\lambda_1,\lambda_2\ldots,)$ we associate, as usual, a Young diagram with $\lambda_1$ boxes in the first row,  $\lambda_2$ in the second, etc. For example,
\be
 \lambda =(3,3,1)\quad\Leftrightarrow \quad  \yng(3,3,1)
 \ee
 Partitions of the form $(k,\underbrace{1,1,\ldots,1}_j)\equiv(k,1^j)$ are  called {\it hooks}.
Partitions may equivalently be expressed in Frobenius notation as
 \be
 \lambda \sim (a_1, \dots , a_k | b_1, \dots , b_k),
 \ee
where  $k$ is the number of diagonal boxes in the Young diagram,  called the rank of the partition and
the $a_i$'s and $b_i$'s are, respectively, the number of boxes to the right of and below the diagonal ones. The Giambelli formula
\be
s_{(a_1, \dots , a_k | b_1, \dots, b_k)} = \det\left(s_{(a_i | b_j)}\right)
\ee
expresses the Schur function corresponding to an arbitrary partition as a determinant whose entries are the Schur functions corresponding to hook diagrams only.  Sato also used Giambelli's formula in the decomposition of the coefficients of the expansion (\ref{tauexpan}), since the same determinantal relations are valid, when expressed for the Pl\"ucker coordinates of the image of any element $w$ of the Grassmannian (i.e. for any completely decomposable element of $\Lambda\HH$), expressing them in terms those for hook partitions (see eq.~(\ref{dethookplucker}). This amounts to an explicit solution of the Pl\"ucker relations, valid on the affine neighborhood corresponding to the ``big cell''
Using eq.~(\ref{plucker_tauderiv}), the resulting relations appear as partial differential equations, for which the tau-function plays the role of generating function.

Such expansions are valid for any tau-function $\tau_w(\tb)$, but here we will mainly consider  $\tau$-functions $\tau(\eb, \tb)$ associated to the KND data on an algebraic curve $X$ of genus $g$, and compute the coefficients in the Schur function expansion for this case.

Let the curve $X$ be equipped with a canonical homology bases $(\mathfrak{a}_1,\ldots,\mathfrak{a}_g;\mathfrak{b}_1,\ldots,\mathfrak{b}_g)$, and corresponding polygonization, and choose a  basis of  holomorphic differentials $\ub=(u_1,\ldots, u_g )^T$
 ordered according to the degree of their vanishing at the the Weierstrass point $p_{\infty}$ at infinity,
$n_g+1,\ldots, n_1+1$, where $\mathfrak{W}=(n_1,\ldots,n_g)$ is the Weierstrass gap sequence at $p_{\infty}$ (see section 3.2).

Denote by
\be
\grA=\left(\oint_{\mathfrak{a_j}} u_i\right)_{i,j=1,\ldots, g} , \quad
\grB=\left(\oint_{\mathfrak{b_j}} u_i\right)_{i,j=1,\ldots, g}\label{abmat}
\ee
the matrices of periods. The Jacobian of the curve is then $\mathrm{Jac}(X)= \mathbb{C}^g/ \grA \oplus \grB$.

We will refer to $(\grA,\grB)$ as the first period matrices or Riemann period matrices. The second period matrices, $(\grS,\grT)$ are similarly formed from the periods of meromorphic differentials
$\rb=(r_1,\ldots, r_g )^T$ with poles only at  the Weierstrass point $p_{\infty}$,
of orders $n_g+1,\ldots, n_1+1$, respectively.

\be
\grS=-\left(\oint_{\mathfrak{a_j}} r_i\right)_{i,j=1,\ldots, g} , \quad
\grT=-\left(\oint_{\mathfrak{b_j}} r_i\right)_{i,j=1,\ldots, g}
\ee
normalized by the condition (generalized Legendre equation)
\be
\left(  \begin{array}{cc}  \grA & \grB\\
                           \grS & \grT  \end{array} \right) \Jb \left(  \begin{array}{cc}  \grA & \grB\\
                           \grS & \grT  \end{array} \right)^T=-2\imath\pi \Jb, \quad \Jb=\left( \begin{array}{cc} \mathbf{0}_g&-\mathbf{1}_g\\  \mathbf{1}_g&\mathbf{0}_g   \end{array}
                           \right).\ee
 The matrix
\be
\boldsymbol{\varkappa}:=  \grA^{-1}\grS \label{varkappa}
\ee
is necessarily symmetric, $\boldsymbol{\varkappa}^T=\boldsymbol{\varkappa}$, and
\be
\grS=\boldsymbol{\varkappa}\grA,\qquad \grT=\boldsymbol{\varkappa}\grB-\frac{\imath\pi}{2} {\grA^{-1}}^T.
\label{stmat}
\ee
\br
 In the basis $(\ub,\rb)$ the meromorphic differentials $\rb$ are not uniquely defined, only up to the addition of a holomorphic differential. Therefore the matrix $\boldsymbol{\varkappa}$ is only defined up to the addition of an arbitrary symmetric matrix. Nevertheless for our purposes, it is sufficient to choose a specific representative of the class of differentials $\rb$, and this will be specified in each case treated in detail in the examples. \label{ubrb}
\er

Following Baker \cite{ba07}, we associate to the curve $X$ the fundamental bi-differential $\Omega(p, q)$, which is the unique symmetric meromorphic $2$-form on $X\times X$, with a second order pole
on the diagonal $p=q$, and elsewhere holomorphic in each variable. Locally expressed, this
has the form
\bea
\Omega(p,q)=\frac{\mathrm{d}\xi(p)\mathrm{d}\xi(q)}{(\xi(p)-\xi(q))^2 }
+\sum_{i,j=0}^{\infty} \mu_{ij}(p_0) \xi(p)^i\xi(q)^j \mathrm{d}\xi(p)\mathrm{d}\xi(q),
\eea
where and $\xi(p),\xi(q)$ are local
coordinates in the vicinity of a base point $p_0$, $\xi(p_0)=0$  and the coefficients $\mu_{ij}(p_0)$ are symmetric in the $i,j$-indices, $\mu_{ij}(p_0)=\mu_{ji}(p_0)$.
 The bi-form $\Omega(p, q)$ is normalized by the conditions
\be
\oint_{\mathfrak{a}_j} \Omega(p, q)  =0,\quad   j=1,\ldots,g. \label{normalization}
\ee

Usually $\Omega(p,q)$ is realized as the second logarithmic derivative of the
prime-form or theta-function \cite{fay73}. But in our development we
use an alternative representation of $\Omega(p,q)$ in an {\em
algebraic form} that goes back to Weierstrass and Klein, as described by
Baker in \cite{ba97}
\be
  \Omega(p,q)=\frac{\mathcal{F}(p,q)}{P_y(p) P_w(q)(x-z)^2}
  \mathrm{d}x\mathrm{d}z+
  2\ub(p)^T\varkappa \ub(q), \label{omegaQS}
\ee
where $p=(x,y)$, $q=(z,w)$, the function $\mathcal{F}(p,q)=\mathcal{F}((x,y),(z,w))$
is a polynomial in its arguments, with coefficients depending on the parameters
defining the curve $X$ explicitly, as a planar model  given by the polynomial equation
\be
P(x,y)=0,
\ee
$\ub$ is the $g$-component vector whose entries are the holomorphic differentials
and $\varkappa$ is a symmetric matrix (\ref{varkappa}) thereby providing the normalization (\ref{normalization})  of
$\Omega(p,q)$.  We will refer to the first term on the right of
(\ref{omegaQS}), which involves the polynomial $\mathcal{F}(p,q)$, as
the algebraic part\footnote{The factor $2$ in the normalization makes the case $g=1$
agree with the usual one in the Weierstrass theory of elliptic functions}, denoting it as $\Omega^{\rm{alg}}(p,q)$.\footnote{The representation  (\ref{omegaQS}) was further developed by Buchstaber, Leykin and one of the authors \cite{bel97b} and more recently by Nakayashiki
\cite{nak08}.}  In the vicinity of the base point $p_{0}$, $\xi(p_{0})=0$ the form  $\Omega^{\rm{alg}}(p,q)$ is expanded in a power series as
\bea
\Omega^{\rm{alg}}(p,q)=\frac{\mathrm{d}\xi(p)\mathrm{d}\xi(q)}{(\xi(p)-\xi(q))^2 }
+\sum_{i,j=0}^{\infty}   \mu^{\rm{alg}}_{ij}(p_{0}) \xi(p)^i\xi(q)^j   \mathrm{d}\xi(p)\mathrm{d}\xi(q),
\eea
where the quantities $\mu^{\rm{alg}}_{ij}(p_{0})$ are algebraic functions of $\xi(p_0)$ and  the coefficients
of the curve.  The transcendental part $\Omega^{\rm{trans}}(p,q)$ of $\Omega(p,q)$ is holomorphic and its expansion in the vicinity of the base point $p_{0}$ has the form
\bea
\Omega^{\rm{trans}}(p,q)&\equiv &  2\ub(p)^T\varkappa \ub(q)\nonumber  \\ &=&
\sum_{i,j=0}^{\infty}   \mu^{\rm{trans}}_{ij}(p_{0}) \xi(p)^i\xi(q)^j   \mathrm{d}\xi(p)\mathrm{d}\xi(q).
\eea
Therefore
\be
\Omega(p,q)=\Omega^{\rm{alg}}(p,q)+\Omega^{\rm{trans}}(p,q)
\ee
 and
 \be
 \mu_{ij}(p_0)=\mu_{ij}^{\rm{alg}}(p_{0}) + \mu_{ij}^{\rm{trans}}(p_{0})
 \ee
 for all $p_0$ and $i,j \in \mathbb{Z}$.

The algebraic representation of the fundamental differential, as
described above, lies behind the definition of the multivariate
sigma function in terms of the mutivariate (Riemann) theta-functions $\theta$. It differs from $\theta$
by an exponential factor and a modular factor $C$:
\begin{equation}
  \sigma(\vb) = C\,\mathrm{exp} \left\{
    \frac12 \vb^T \boldsymbol{\varkappa}\vb  \right\}
  \theta\left(\grA^{-1}\vb;\tau \right), \label{sigma}
\end{equation}
where $\boldsymbol{\varkappa}$ and $\grA$ are defined in (\ref{varkappa}) and (\ref{abmat}).
These modifications make $\sigma(\boldsymbol{v})$ invariant with
respect to the action of the symplectic group, so that for any
$\gamma\in{\mathrm Sp}(2g, \mathbb{Z})$, we have:
\be
\sigma(\vb;\gamma\circ \mathfrak{M})=\sigma(\vb;\mathfrak{M}),
\ee
where $\mathfrak{M}$ is the set of periods of the curve $X$.
The multivariate sigma-function is the natural generalization of the
Weierstrass sigma function to algebraic curves of higher genera.

\br In his lectures, \cite{w-lect}, Weierstrass defined  the sigma-function in terms of
a series with coefficients given recursively, a key point of the
Weierstrass theory of elliptic functions. A generalization of this
result to genus two curves was begun by Baker \cite{ba07} and
recently completed by Buchstaber and Leykin \cite{bl05}, who obtained
recurrence relations between coefficients of the sigma-series in
closed form. Buchstaber and Leykin also recently found an
operator algebra that annihilates the sigma-function of higher
genera $(n,s)$-curves \cite{bl08}. Finding a suitable recursive definition of the
higher genera sigma-functions along the lines of  \cite{bl08}  remains a challenging problem.
\er

In this paper we study the relation between the multivariate sigma function
and the Sato $\tau$-function \cite{sato80,sato81,sato82} for the case of quasi-periodic solutions
associated to  KND data on an algebraic curve. Thus, we mainly consider this class of ``algebro-geometric
$\tau$-function''  solutions. These are essentially the same as those
studied by Fay \cite{fay83,fay89} in terms of $\theta$-functions. In terms of $\sigma$ functions
such $\tau$-functions may be expressed (see Proposition \ref{tau_sigma_prop})\footnote{Recently A. Nakayashiki \cite{nak09} has independently suggested a similar expression for the algebro-geometric tau functions in terms of multivariate $\sigma$-functions and studied properties of the sigma-series.}
\bea
  \frac{ \tau(\eb, \tb)}
  {\tau(\eb,  \boldsymbol{0})}= \frac{
    \sigma\left(\sum_{k=1}^{\infty} \grA\Ub_k(p_\infty) t_k+
      \grA\eb\right)} {\sigma(\grA\eb)}\mathrm{exp}
  \left\{ \frac12 \sum_{k,l=0}^{\infty}
    \mu^{\mathrm{alg}}_{kl}(p_\infty)t_k t_l \right\}.
\label{tausigma}
\eea
Here, as above,  $\grA$ is the period matrix of holomorphic differentials, $\Ub_k(p_{\infty}), k=1,2,\ldots$ are ``winding vectors" i.e. $\mathfrak{b}$ -periods of normalized second kind differentials with poles of order $k+1$ at the point $p_{\infty}$, and
$\eb$ is an arbitrary point of the Jacobian variety $\mathrm{Jac}(X)$.

The paper is organized as follows. In section 2, we review the geometric formulation
of the Sato-Segal-Wilson $\tau$-function in terms of Hilbert space Grassmannians. The interpretation of the coefficients of the Schur function expansion as Pl\"ucker coordinates is derived (Proposition \ref{satoprop}),
as well as their expression in terms of the affine coordinates on the big cell, and hook partitions
(Corollary \ref{det_affine_expansion}).

In section 3, we restrict to the special case of $\tau$-functions associated to  KND data on
an algebraic curve. The explicit formula for such $\tau$-functions in terms of Riemann $\theta$-functions
is given in eq.~(\ref{alg_curve_tauQ}). The affine coordinates determining the Schur function expansion are computed explicitly (eq.~(\ref{affine_quadratic})) in terms of directional derivatives of the Riemann
$\theta$-function along the flow directions. In subsection 3.2, we review the Weierstrass gap theorem and introduce the normalized symmetric bi-differential $\Omega(p, q)$. This allows us to make explicit the
splitting into ``algebraic'' and ``transcendental'' parts (Corollary \ref{cor31}) of the infinite quadratic form
$Q$ appearing in the exponential term in  eq.~(\ref{alg_curve_tauQ}) and the formula eq.( \ref{affine_quadratic}) determining the affine coordinates. The equivalent expression for the $\tau$-function in terms of the $\sigma$-function, which displays more clearly its modular transformation properties is derived in Proposition \ref{tau_sigma_prop},
eq.~(\ref{tau_sigmae1}). From this, it is shown how to compute the affine coordinate matrix elements explicitly as polynomials in terms of the Kleinian functions $\{\zeta_i, \wp_{ij}\}, $ that
generalize  the Weierstrass $\zeta$- and $\wp$-functions to curves of arbitrary genus.

In section 4, we consider several examples for which the affine coordinates are explicitly calculated and use the Pl\"ucker relations to derive  identities relating Kleinian $\zeta$ and $\wp$ functions of different degrees. The case of hyperelliptic curves is treated in detail, with explicit formulae computed for the case of genus $g=2$.  In Proposition \ref{3_2_identities_prop} we derive algebraic relations between the  Kleinian $\wp$-functions of different orders for this case. A further example based on a planar model of a class of trigonal curves is considered, and explicit expressions for the lowest affine coordinate matrix elements are computed in terms of the Kleinian $\zeta$ and $\wp$-functions for this case, as well as identities relating $\wp$-functions of different orders that follow again from the Pl\"ucker relations.

\section{Background. Sato-Segal-Wilson $\tau$-function}

The following is a brief summary of the Sato \cite{sato82} and Segal and Wilson \cite{sw85} approach to KP $\tau$-functions. The first is essentially algebraic in nature, the second functional analytic, but we combine  elements of both. For more precise definitions of the main ingredients (Hilbert space Grassmannians, infinite transformation groups, determinant line bundles,
Fermionic Fock space,  the Pl\"ucker map, etc.), the reader is referred to these two sources, which agree in the geometric framework but not in analytic details. The introductory summary given here is intended to be as simple and self-contained as possible,  and applicable to the cases at hand. Functional analytic details of the general formulation are either omitted or referred to \cite{sw85}, and the ingredients are made as much as  possible to appear  like the finite dimensional case.

\subsection {Hilbert space Grassmannian and Pl\"ucker coordinates}

  Following \cite{sw85}, we start with the Hilbert space
  \be
  \HH:= L^2(S^1) = \HH_+ \oplus \HH_-
  \ee
  of square integrable complex valued functions  $f$ on the unit circle $\{z= e^{i\phi}\}$ in the complex plane, which may be split as a sum
  \be
f = f_+ + f_-, \quad f_\pm \in \HH_{\pm},
\ee
where $f_{\pm} \in \HH_{\pm}$ are the positive and negative parts  of the Fourier series.

 Equivalently, $\HH_+$ may be interpreted as the space of holomorphic functions on the interior unit disk and  $\HH_-$  as the space of holomorhic functions on the exterior, vanishing at $z=\infty$, with orthonormal bases  provided by the monomials in $z$
  \be
  \HH_+ =\span\{e_j:=z^{-j-1}\}_{j=-1,-2 \dots, }, \qquad
  \HH_- =\span\{e_j:=z^{-j-1}\}_{j=0,1,2  \dots }.
  \ee
  \br The convention of labeling the basis vectors $e_j$ so $\mathcal{H}_+$ is the span of those having negative indexes $j$ is chosen so that under the Pl\"ucker map (see below) $\mathcal{H}_+$ is taken into  $|0\rangle=e_{-1}\wedge e_{-2}\wedge \ldots$ which is the ``Dirac sea", in which all negative ``energy" stated are occupied.
  \er

  We denote by $\Gr_{\HH_+}(\HH)$ the Hilbert space Grassmannian whose points are
  closed subspaces $w\subset \HH$ that are commensurable with $\HH_+$ in the sense that
  orthogonal projection  \hbox{$\pi^\perp: w \ra \HH_+$ to $\HH_+$}
  along $\HH_-$  is a Fredholm map with index zero and orthogonal projection
 \hbox{$ \pi^\perp: w \ra\HH_-$} to $\HH_-$ along $\HH_+$ is Hilbert-Schmidt.
   (In \cite{sw85} this is called the {\it zero virtual dimension  } component of  the full Hilbert space Grassmannian.)

  Let
  \bea
  w &\&= \span\{w_j\}_{j\in \Nb} \nonumber \\
  w_j &\&= \sum_{i\in \Zb}w_{ij}e_i.
  \eea
   Relative to the monomial basis $\{e_j\}$, the frame $\{w_0, w_1, \dots\}$ may be represented as an infinite matrix $W$ with components $\{w_{ij}\}_{i\in \Zb, j\in \Nb}$  whose $j$th column vector $W_j$ has components $\{w_{ij}\}_{i\in \Zb}$
   \be
   W = (W_0, W_1, \dots ) = \begin{array}{cc}
   \left(\begin{array}{c}\\ W_+\\  \\ \\ \\W_-\\\\\end{array}\right)&
   \begin{array}{c}  {}_{-\infty}\\ \vdots\\ {}_{-1}\\ \\ {}_{0}\\ \vdots\\{}_{ +\infty} \end{array} \end{array}
   \label{Wplusminus}
   \ee
   where the rows are labeled by the integers, increasing downward
    with the $0$-th row  at the top  of the $W_-$ block, and the columns are labeled by the
    non-negative integers, starting with $0$ on the left and increasing to the right.
    \br Note that the column labeling corresponds to the monomials degrees in $\HH_+$ whereas the row labeling corresponds to the basis elements $\{e_j\}$
    \er
    Here $W_-$ may be viewed as representing a map $w_-: \HH_+ \ra \HH_-$
   and  $W_+$ a map $w_+: \HH_+ \ra \HH_+$ such that  $w$ is the image of their sum
   \be
   w = (w_-  + w_+)(\HH_+).
   \ee

   The Grassmannian $\Gr_{\HH_+}(\HH)$ may be  interpreted
   as the quotient of the bundle $\Fr_{\HH_+}(\HH)$ of admissible  frames by the right action
   of the group   $\grGl(\HH_+)$ of  linear changes of basis.  (See \cite{sw85} for the precise definition, which  requires elements of $\grGl(\HH_+)$ to have a non-vanishing, finite determinant). Thus, $\Fr_{\HH_+}(\HH)$ may be viewed, similarly to
   the finite dimensional case, as a principal  $\grGl(\HH_+)$ bundle over $\Gr_{\HH_+}(\HH)$),
   to which we may associate, through the determinant representation, a line bundle
   $\Det \ra \Gr_{\HH_+}(\HH)$ and its dual  $\Det^* \ra \Gr_{\HH_+}(\HH)$. A holomorphic
   section of the latter is determined by each partition $\lambda =(\lambda_1 \ge \lambda_2 \ge
   \dots \ge\lambda_{ \ell(\lambda)} > 0)$,    $\lambda_i\in \Nb$, where $\ell(\lambda)$ denotes the length,   in a way that mimics the finite dimensional case. Extending the set of parts
   $\{\lambda_j\}_{j=1,\dots \ell(\lambda)}$ in the usual way \cite{mac90}  to an infinite sequence
   $\{\lambda_j\}_{j=1,\dots \infty}$ by defining,
    \be
   \lambda_j =0  \quad  {\rm if \quad} j > \ell(\lambda) ,
   \ee
the determinant  $\det(W_\lambda)$ of the submatrix $W_\lambda$ of $W$ consisting of the rows $ \{{\lambda_i  -i}\}_{i\in \Nb^+}$ defines a holomorphic section
$\sigma:\Gr_{\HH_+}(\HH) \ra \Det^*$
 of the bundle $\Det^* \ra \Gr_{\HH_+}(\HH)$ and these span the space of admissible sections. (Again, for the full analytic details required to define the class of admissible sections, see \cite{sw85}.)

   As in the finite dimensional case, the space of such holomorphic sections may be identified
 with a certain subspace $\FF_0 \subset \FF$ of the exterior space $\FF :=\Lambda \HH$, the ``zero charge sector'' of $\FF$,  with the latter interpreted as the full ``Fermionic Fock space''.
  Let   $\FF_0$ be the span of the exterior elements
  \be
  \label{pluckerbasis}
|\lambda \rangle:=e_{l_1} \wedge   e_{l_2} \wedge  e_{l_3} \wedge \cdots \ ,
  \ee
  where, for any partition $\lambda=(\lambda_1,\lambda_2,\ldots)$, the $l_i$'s are the ``particle coordinates", \be l_{i}=\lambda_i-i.\label{particle}\ee
  These form an orthonormal basis with respect to the inner product induced on
  $\FF_0 \subset \Lambda \HH$ by the one on $\HH$.

  For each partition $\lambda$, we may also define $\HH_\lambda \in \Gr_{\HH_+}(\HH)$ as
\be
\HH_\lambda =\span\{e_{l_i}\}_{i\in \Nb^+}
\ee
In particular, the element  $\HH_0$ corresponding to the trivial partition $\lambda = 0$
is $\HH_+$.
  We define the Pl\"ucker  map $\hat{\grP} : \mathrm{Fr}_{\HH_+}(\HH) \ra \FF_0 $ in the natural way
  \bea
  \hat{\grP} : \mathrm{Fr}_{\HH_+}(\HH) &\&\ra \FF_0 \\
 \hat{ \grP}: \{w_0, w_1, \dots \} &\& \mt  w_0 \wedge w_1 \wedge w_2  \wedge \cdots
 \label{pluckerhat}
 \eea
Changing the frame $\{w_0, w_1, \dots \}$ spanning the subspace $w\subset \HH $  by
application (on the right) of an element $ g \in \grGl(\HH_+)$  just changes the image under $\hat{\grP}$ by the nonzero multiplicative factor $\det(g)$. Therefore the Pl\"ucker map (\ref{pluckerhat}) on the frame bundle projects to a map embedding $\Gr_{\HH_+}(\HH)$ into the projectivization $\Pb(\FF_0)$ of $\FF_0$
  \bea
\grP : \Gr_{\HH_+}(\HH) &\&\ra \Pb(\FF_0)\nonumber \\
 \grP: \{w_0, w_1, \dots \} &\& \mt  [w_0 \wedge w_1 \wedge w_2  \wedge \cdots] ,
  \label{plucker}
 \eea
 where $[|v \rangle]$ denotes the projective equivalence class of $|v \rangle \  \in \FF_0$.
 In particular,
 \be
 \grP (\HH_\lambda) = [|\lambda \rangle] .
 \ee
 The image of $\Gr_{\HH_+} (\HH)$ under $\grP$ is the orbit of $\HH_+$ under the identity component of the general linear group $\grGl(\HH)$ (again,
 suitably defined, as in \cite{sato82} or \cite{sw85}).

 The Pl\"ucker  coordinates \{ $\pi_\lambda(| v \rangle)$\} of an element $| v\rangle \in \FF_0$ are just its
  components relative to the orthonormal basis $\{|\lambda \rangle\}$:
 \bea
 \pi_\lambda(|v\rangle) &\& = \ \langle\lambda  | v\rangle \cr
 |v\rangle &\& = \sum_{\lambda}\pi_\lambda | \lambda \rangle.
  \eea
Applying  the Pl\"ucker map $\hat{\grP}$ to an  element $\{w_0, w_1, \dots \} \in \mathrm{Fr}_{\HH_+}(\HH)$
spanning  $w\in \Gr_{\HH_+(\HH)}$,  the Pl\"ucker coordinates  of its image are the homogeneous
coordinates of  $\grP(w)$ under the  Pl\"ucker map
(\ref{plucker})
\be
\pi_\lambda(w):= \pi_\lambda(\grP(w)).
\ee
It follows from the above that
\be
\pi_\lambda(w) = \det(W_\lambda),
\ee
and hence the basis of holomorphic sections  $H^0( \Gr_{\HH_+}(\HH), \Det^*)$ of $\Det^*$
defined by the partitions $\lambda$ correspond precisely to the Pl\"ucker coordinates.

The Pl\"ucker relations are the infinite set of quadratic relations that determine the image of
$\Gr_{\HH_+}(\HH)$ under the Pl\"ucker map. They follow from the fact that for any
$w\in \Gr_{\HH_+}(\HH)$, this image $\grP(w)$ is a decomposable element of $\mathcal{F}_0$.
  The Pl\"ucker coordinates of $\pi_\lambda(w)$ are therefore not independent;
it is possible to express them on an open dense affine subvariety as finite determinants  in terms of a much smaller subset consisting,  e.g.,  of those Pl\"ucker coordinates corresponding to hook partitions. It is convenient to use  Frobenius notation $(a_1 a_2 \cdots a_r | b_1 b_2 \cdots b_r)$ for a partition,
\cite{mac90}, where $(a_i, b_i)$ are the number of elements to the right and below the $i$th diagonal element of the Young diagram, for $i=1 \dots r$. A hook partition  $\lambda = (a+1, 1^b)$ is  one for which $r=1$, and hence in Frobenius notation is expressed $(a\  | \ b)$.

To see how to express the Pl\"ucker coordinate  $\pi_{(a_1 a_2 \cdots a_r | b_1 b_2 \cdots b_r)}(w)$
corresponding to an arbitrary partition in terms of those for the hook partitions
$\{(a_i \ | \ b_j)\}_{1\le i,j \le r}$,  it is easiest to assume that $w$ is in the ``big cell'', in which the map
$w_+: \mathcal{H}_+\ra \HH_+$ is invertible. The infinite matrix $W_+$ in (\ref{Wplusminus}) is therefore also
invertible, and we may define affine coordinates as the matrix  entries of
\be
A := W_-W_+^{-1} .
\label{affinecoords}
\ee
By convention, we will let the indices $(a, b)$ range over the non-negative integers,
and therefore the componentwise interpretation of (\ref{affinecoords}) is
\be
A_{ab}:= (W_-W_+^{-1})_{ab},   \qquad a,b \in \mathbb{N}.
\ee
The homogeneous coordinates in this basis have the form
\be
WW_+^{-1} = \left(\begin{array}{c} \Ib \\ A\end{array}\right),
\ee
where $\Ib$ is the semi-infinite  identity matrix, $I_{ij}=\delta_{-i-1,j}$ which,  in view of the labeling convention has $\{ij\}$-th entry $i\leq -1, \;j\geq 0 $  and the labeling convention for the matrix $A$ in the lower block is that  the pair of indices $(ab)$ start with $(0,0)$ in the upper left-hand corner, and increase downward and to the right. This allows us to express all the Pl\"ucker
coordinates as finite determinants in terms of the affine coordinates on the big cell.
\bp \label{pluckercoordinate}
The Pl\"ucker coordinate $\pi_{(a_1 a_2 \cdots a_r | b_1 b_2 \cdots b_r)}(w)$ corresponding to the partition $(a_1 a_2 \cdots a_r | b_1 b_2 \cdots b_r)$
 is
\begin{eqnarray}
\pi_{(a_1 a_2 \cdots a_r | b_1 b_2 \cdots b_r)}(w)&=&  \det (W_{(a_1 a_2 \cdots a_r | b_1 b_2 \cdots b_r)})\nonumber\\&=& (-1)^{\sum_{k=1}^r b_k}  \det \left(A_{a_i, b_j} |_{1\le i,j \le r}\right) \mathrm{det}(W_+)\ .
  \label{pluckerfrobenius}
\end{eqnarray}
\ep

In particular, we may consider the case of hook partitions, which, according to (\ref{pluckerfrobenius}),
coincide within a sign with the components of the affine coordinate matrix $A$, allowing all other
Pl\"ucker coordinates to be expressed as finite determinants in terms of these.
\bc
\label{dethookplucker_coroll}
The Pl\"ucker coordinate corresponding to a hook partition $(a | b )$ is,
within a sign,  the $(a,b)$ affine coordinate,
\be
\pi_{(a | b)} (w)  = (-1)^b A_{ab},
\label{hookplucker}
 \ee
and hence
\be
\pi_{(a_1 a_2 \cdots a_r | b_1 b_2 \cdots b_r)}(w)=\mathrm{det}\left( \left.\pi_{(a_i | b_j)} (w)\right|_{1\leq i,j \leq r} \right).
\label{dethookplucker}
\ee
\ec

\subsection {Abelian flow group $\Gamma_+$ and the KP $\tau$-function}

We now introduce the abelian subgroup $\Gamma_+ \subset \grGl(\HH)$ consisting of
nonvanishing  elements of $\HH_+$ normalized to equal $1$ at the origin $z=0$
\bea
\Gamma_+&\& := \{ \gamma(\tb) :=e^{\sum_{i=1}^\infty t_i z^i}\} \cr
\tb &\& := (t_1, t_2, \dots \},
\eea
acting on $\HH$  by multiplication
\bea
\Gamma_+ \times \HH &\&\ \ra \  \HH \cr
(\gamma, f) &\&\ \mt \ \gamma f .
\label{Gammaflow}
\eea
This induces an action on  the Grassmannian
\bea
\Gamma_+ \times Gr_{\HH_+}(\HH) &\&\ \ra \ Gr_{\HH_+}(\HH) \cr
(\gamma(\tb), w ) &\&\ \mt \ \gamma(\tb) w
\eea
which lifts to one on the bundle $\Det^* \ra \Gr_{\HH_+}(\HH)$,
and determines an action on the space of holomorphic sections $H^0(\Det^*, \Gr_{\HH_+}(\HH))$
\bea
\Gamma_+ \times H^0(\Det^*, \Gr_{\HH_+}(\HH)&\&\ \ra \ H^0(\Det^*, \Gr_{\HH_+}(\HH) \cr
(\gamma(\tb), \sigma) &\&\ \mt \ \tilde{\gamma}(\tb) \sigma := \sigma \circ \gamma^{-1}(\tb) \cr
 \tilde{\gamma}(\tb) \sigma (w) &\&:=  \tilde{\gamma}(\tb) \sigma (\gamma^{-1}(\tb) w)
\eea
The latter coincides, within  normalization, with
the induced action  on $\FF_0\subset \Lambda \HH$. Let
\be
w(\tb) = \gamma(\tb)(w)
\ee
be the image of $w\in \Gr_{\HH_+}(\HH)$ under the action of the group element $\gamma(\tb)$
and let $W(\tb)$ be its matrix of homogeneous coordinates relative to the standard basis
of monomials $\{e_j\}_{j\in \Zb}$.
Then
\be
\pi_\lambda(w(\tb)) = \det (W_\lambda (\tb))
\ee
is the Pl\"ucker coordinate of $w(\tb)$ corresponding to the partition $\lambda$.
The KP $\tau$-function is defined to be the Pl\"ucker coordinate corresponding
to the trivial partition $\lambda = 0$
\be
\tau_w(\tb) : = \pi_0(w(\tb)) = \det(W_+(\tb)).
\ee

Since, as will be seen in the
next subsection, all other Pl\"ucker coordinates of $w(\tb)$ may be determined from
$\tau_w(\tb)$ by applying  constant coefficient differential
 operators in the $\tb$ variables,
defined in terms of the Schur functions, the Pl\"ucker relations for $\grP(w(\tb))$ may be expressed as an infinite system of bilinear differential relations satisfied by $\tau_w(\tb)$,  the Hirota relations \cite{sato82}, which are equivalent to the hierarchy of KP flow equations.

\subsection {Schur function expansions}

  Recall that if we view the flow parameters $\tb =(t_1, t_2,  \dots )$ as power
  sums in terms of a set $\{x_1, \dots , x_N\}$ of $N$ auxiliary variables
  \be
  t_i = {1\over i} \sum_{a=1}^N x_a^i ,
  \label{powersums}
  \ee
  the Schur function $s_\lambda$, which is the irreducible character of the
  tensor representation of $\grGl(N)$ with symmetry type  corresponding to
  partition $\lambda$, is given by the Jacobi-Trudi formula \cite{mac90}
  \be
  s_\lambda(\tb) = \det (h_{\lambda_i -i +j})\vert_{1\le i,j \le n}
  \label{JT}
  \ee
  for any $n\ge \ell(\lambda)$, where $\{h_j(\tb)\}_{j=1, \dots \infty}$ are the complete symmetric functions
  defined by the generating function formula
  \be
  e^{\sum_{i=1}^\infty t_iz^i} =  \sum_{j=0}^\infty h_j(\tb) z^j.
  \label{hjgener}
  \ee
 We have $h_0(\tb)=1$ and it is understood in (\ref{JT}) that $h_j(\tb):=0$ for $j < 0$.

    Given any function $f(\tb)$ that admits a Taylor expansion  in the flow variables about
   the  origin $\0b :=(0,0,\dots)$,
   \be
      f(\tb) =( e^{{\sum_{i=1}^\infty t_i {\partial\over \partial\tilde{t}_i}}} f(\tilde{\tb})) |_{\tilde{\tb}=\0b},
      \ee
      where $\tilde{\tb} = (\tilde{t}_1, \tilde{t}_2, \dots )$,
   we may use the Schur functions as a basis and express the series as
   \be
   f(\tb) = \sum_{\lambda} f_\lambda s_\lambda(\tb) .
   \ee
   (This determines the ``Bose-Fermi equivalence'', which associates an element
   $\sum_{\lambda} f_\lambda | \lambda \rangle$ of the Fermi Fock space $\FF_0$ to the Bose-Fock space
   element $f$, viewed as a symmetric function of an underlying infinite series of
   parameters $\{x_a\}_{a=1, \dots \infty}$ related by (\ref{powersums}), in the inductive limit, to the flow parameters $\tb$.)
   Using the Cauchy-Littlewood identity \cite{mac90}
  \be
   e^{\sum_{i=1}^{\infty} i t_i\tilde{t}_i} =   \sum_{\lambda} s_\lambda (\tb) s_\lambda (\tilde{\tb})
    \ee
    in the form
   \be
   e^{\sum_{i=1}^{\infty} t_i {\partial \over \partial t_i} }   =  \sum_{\lambda} s_\lambda (\tb) s_\lambda (\partial_{\tb}),
    \ee
   where
   \be
   \partial_{\tb} :=\left\{{1\over i}{\partial \over \partial t_i}\right\}_{i=1,2, \dots},
   \ee
   we obtain
   \be
   f_\lambda = s_\lambda (\partial_{\tb}) (f (\tb)) \vert_{\tb=\0b} .
  \label{schurderiv}
   \ee

   For the  tau function $\tau_w(\tb)$, the coefficients in this expansion coincide
   with the Pl\"ucker coordinates $\pi_\lambda(w)$ of the initial point $w \in Gr_{\HH_+}(\HH)$.
   \bp
    [\bf Sato \cite{sato82}] \label{satoprop}
   The Schur function expansion of  $\tau_w(\tb)$ is
   \be
   \tau_w(\tb) =  \sum_\lambda  \pi_\lambda(w) s_\lambda(\tb).
   \label{shurtauseries}
   \ee
   The Pl\"ucker coordinates are therefore given by
   \be
    \pi_\lambda(w) =s_\lambda (\partial_{\tb}) \left( \tau_w(\tb)\right) \vert_{\tb=\0b}.
    \label{schurderivplucker}
   \ee
   \ep
   \noindent{\bf Proof}. Using  formula (\ref{hjgener}), it is easy to see that the matrix representation
 of the action (\ref{Gammaflow}) is given by:
\be
 W(\tb) = \pmatrix{ H_{++}(\tb) & H_{+-}(\tb) \cr 0 & H^T_{--}(\tb)} \pmatrix{ W_+ \cr W_-},
 \ee
 where
 \bea
 H_{++}(\tb)&\&:= \pmatrix{    \ddots &\ddots &\vdots &  \vdots \cr
         \ddots &1 &h_1 &  h_2 \cr
      \cdots &0 &1  & h_1 \cr
      \cdots &0 & 0 &1 }   \cr &\&\cr
  H_{+-}(\tb) := \pmatrix{   \vdots &  &  &               \cr
  h_3 & \vdots & &            \cr
  h_2 & h_3 & \vdots &                \cr
    h_1 & h_2 & h_3 & \cdots},  &\&
    \qquad
     H_{--}(\tb) := \pmatrix{ 1& h_1 & h_2 & h_3 & \cdots\cr
    0 & 1  & h_1 & h_2 & \cdots \cr
      0 &  0 &1 & h_1 & \cdots \cr
      \vdots &  \vdots  & \vdots & \ddots & \ddots
     }.
 \eea
 Letting
 \be
 H(\tb):= \pmatrix{  H_{++}(\tb)&  H_{+-}(\tb) },
 \ee
we have
 \be
 \tau_w(\tb) = \det(H(\tb) W) = \sum_\lambda \det(H_\lambda(\tb))\det(W_\lambda)
 = \sum_\lambda \pi_\lambda(H^T(\tb))\pi_\lambda(w) ,
 \ee
  where the second equality is the Cauchy-Binet identity and
  \be
   \pi_\lambda(H(\tb)) := \det (h_{\lambda_i -i +j}(\tb)) |_{1\le i,j \le  \infty} =s_\lambda(\tb)
   \ee
   by (\ref{JT}). We thus obtain the Schur function expansion (\ref{shurtauseries}) of the $\tau$-function.

On the ``big cell'', each $\pi_\lambda(w)$  is determined  through (\ref{pluckerfrobenius}) as a finite determinant in terms of  the affine coordinates $\{A_{ab}\}$ of $w$ which, by (\ref{hookplucker}), coincide within sign with the hook partition Pl\"ucker coordinates. Therefore, we need only apply (\ref{schurderivplucker}) to obtain these
\be
(-1)^b  A_{ab}=  \pi_{(a|b)}(w) =s_{(a|b)} (\partial_{\tb})   \left( \tau_w(\tb) \right) \vert_{\tb=\0b}.
    \label{schurderivhook}
   \ee

Substituting the expression (\ref{pluckerfrobenius}) for the Pl\"ucker coordinates $\pi_{(a_1,\ldots,a_r|b_1,\ldots,b_r)}$ in (\ref{shurtauseries}) we obtain
\bc
\be  \frac{ \tau_{w}(\mathbf{t})}{ \tau_{w}(\mathbf{0})} = \sum_{\lambda}(-1)^{\sum_{k=1}^r b_k} \mathrm{det}\left(\left.{A}_{a_i,b_j}\right|_{1\leq i,j\leq r}\right)s_{\lambda}(\mathbf{t}).  \ee
\ec

A further simplification may be made using the following identity, for which a simple proof follows from the above definitions.
    \bl\label{hooklemma}
   \be
   s_{(a|b)}(\tb) = (-1)^{b}\sum_{j=1}^{b+1} h_{b-j+1}(-\tb) h_{a+j} (\tb).
   \label{hookschurquadratic}
   \ee
   \el
\noindent{\bf Proof}.
By the Jacobi-Trudi formula (\ref{JT}),
\be
 s_{(a|b)}(\tb)  = \det \pmatrix {\hb^T & h \cr \Hb & \kb},
\label{detabmatrix}
 \ee
 where
 \bea
  \fb^T &\&= (h_{a+1}(\tb),   h_{a+2}(\tb) \cdots h_{a+b}(\tb)),
  \quad \quad h:= h_{a+b+1}(\tb),
  \label{bfh}  \\
 \kb&\&:= \pmatrix {h_b(\tb) \cr h_{b-1}(\tb) \cr  \vdots \cr h_1(\tb)}, \quad
 \Hb := \pmatrix{1 & h_1(\tb) & h_2(\tb) & \cdots & \cdots  & h_{b-1}(\tb) \cr
 0 &1 & h_1(\tb) & h_2(\tb) & \cdots  & h_{b-2}(\tb)\cr
0&  0 &\ddots& \ddots&\ddots & \vdots   \cr
\vdots&  \vdots  &   &   &   &\vdots  \cr
0&  0 &\cdots&  &  \cdots & 1  \cr
 }
 \eea
 It follows from the generating function formula (\ref{hjgener}) that the inverse $\Hb^{-1}$
 is given by
 \be
  \Hb^{-1}  := \pmatrix{1 & h_1(-\tb) & h_2(-\tb) & \cdots & \cdots  & h_{b-1}(-\tb) \cr
 0 &1 & h_1(-\tb) & h_2(-\tb) & \cdots  & h_{b-2}(-\tb)\cr
0&  0 &\ddots& \ddots&\ddots & \vdots   \cr
\vdots&  \vdots  &   &   &  & \vdots  \cr
0&  0 &\cdots&  & \cdots & 1  \cr
 }
 \label{Hinverse}
\ee
   and
   \be
   \sum_{j=-a}^b h_{a+j}(-\tb) h_{b-j} (\tb) =\delta_{ab}.
   \label{hjorthog}
   \ee
   The matrix
   \be
 \pmatrix{  \0b^T &  \Hb^{-1} \cr 1 & - h^{-1} \hb^T \Hb^{-1}}
 \ee
 has determinant $(-1)^b$, and its product with the matrix in (\ref{detabmatrix}) is
 \be
 \pmatrix {\hb^T & h \cr \Hb & \kb}  \pmatrix{  \0b^T &  \Hb^{-1} \cr 1 & - h^{-1} \hb^T \Hb^{-1}}
 = \pmatrix{ h & \0b^T \cr
    \kb &  \Ib - h^{-1}\kb \hb^T \Hb^{-1}
    }.
 \ee
 Therefore
 \be
  s_{(a|b)}(\tb) = (-1)^b h \  \det (\Ib - h^{-1}\kb \hb^T \Hb^{-1})
  =(-1)^b (h -  \Hb^{-1} \kb \hb^T).
  \ee
But it follows from (\ref{Hinverse}) and (\ref{hjorthog}) that
   \be
 \Hb^{-1}    \kb= - \pmatrix{h_b(-\tb) \cr h_{b-1}(-\tb)  \cr \vdots \cr h_1(-\tb)},
   \ee
      from which  (\ref{hookschurquadratic}) follows,  in view of the definition (\ref{bfh}) of $h$.
QED.
\hfill \break \noindent

Substituting the identity (\ref{hookschurquadratic}) into (\ref{schurderivhook}) thus gives
\bc  \label{det_affine_expansion}
\be
  A_{ab}= - \sum_{j=0}^b h_{a+j-1}(-\partial_{\tb}) h_{b-j} (\partial_{\tb})  \left( \tau_w(\tb) \right)  \vert_{\tb=0}.
    \label{schurderivbilin}
   \ee
   \ec

     The relations (\ref{pluckerfrobenius}),  (\ref{schurderivplucker})    (\ref{schurderivhook}) determining the Pl\"ucker coordinates of $ w \in \Gr_{\HH_+}(\HH)$ on the ``big cell'' are equivalent to the fact that the formal Baker-Akhiezer function \cite{sato80},\cite{sato81}, \cite{sw85}, defined by the Sato formula \cite{sato82}
  \be
  \psi_w (z, \tb) = e^{{\sum_{i=1}^\infty}t_iz^i} {\tau_w(\tb - [z^{-1}]) \over \tau_w(\tb)},
  \label{sato}
    \ee
where
\be
[z^{-1}]:= \left({1\over z}, {1\over 2 z^2}, {1\over 3 z^3}, \dots \right),
\ee
takes its values in $ w \in \Gr_{\HH_+}(\HH)$ for all values of $\tb$.

Following Sato, we may also introduce the dual Baker function:
\be \Psi^{\ast}_w(z,\tb)= - \frac{\tau_{w}(\tb +[z^{-1}])}{\tau(\tb)}\mathrm{exp}
\left\{  -\sum_{i=1}^{\infty} t_iz^i \right\}. \label{dualbaker} \ee
Then, as shown in \cite{sato80,sato81,sato82}, the KP hierarchy equations may all be expressed in the form of Hirota bilinear equations for the $\tau$-function.

\bt [Hirota bilinear relation \cite{sato80,sato81,sato82}]
\be  \mathrm{Res}_{z=0}\, \Psi^{\ast}_z(z,\tb)  \Psi^{\ast}_z(z,\widetilde{\tb})\equiv 0 ,\label{hirota}\ee
where the $\mathrm{Res}_{z=0}$ just means the coefficient of $z^{-1}$ in the formal Laurent expansion about $z=0$ and the relation is satisfied identically in the infinite set of KP flow variables $\tb=(t_1,t_2,\ldots)$, $\widetilde{\tb}=(\widetilde{t}_1,\widetilde{t}_2,\ldots)$.
\et

\br
In view of (\ref{dualbaker}) , eq.(\ref{hirota}) is equivalently written as
\be
\mathrm{Res}_{z=0}\mathrm{exp}\left\{\sum_{i=1}^{\infty} t_iz^i \right\}\mathrm{exp}\left\{-\sum_{i=1}^{\infty} \widetilde{t}_iz^i \right\}
\tau(\tb-[z^{-1}])\tau(\widetilde{\tb}+[z^{-1}])\equiv 0\label{hirota1}
\ee
identically in $\tb$ and $\widetilde{\tb}$.
\er

\br
It is also shown in \cite{sato80,sato81,sato82} that (\ref{sato}) is simply an expression of the infinite set of Pl\"ucker relations satisfied by the coefficients $\pi_{\lambda}(w)$ appearing in the Schur function expansion (\ref{shurtauseries})
\er

\section{ Algebraic curves}
\subsection {Baker-Akhiezer function and tau function for algebraic curves}

A particularly important class of tau functions  consists of those associated to algebraic curves
\cite{dub81, Kr77, djkm83}.
The relevant data needed to define these are: an algebraic curve $X$ of genus $g$, a positive nonspecial divisor of degree $g$
\be
\DD:= \sum_{i=1}^g p_i,  \quad p_i\in X,
\ee
(or, equivalently, a degree $g$ line positive bundle $\LL \ra X$ in general position satisfying
suitable generic stability conditions), a point ``at infinity'' $p_\infty \in X$ and a local
parameter $\xi =\frac{1}{z}$ defined on a disc
\be
D_\infty := \{p(\zeta), \quad  |\zeta| \le 1, \quad p(0)  = p_\infty\}
\ee
centered at $p_\infty$. The points $p_i$ are assumed to lie in the complement $D_0 := X - D_\infty$.  Identifying $S^1$ with $\partial D_\infty$, the associated element $ w:=w(X,\mathcal{D},p_{\infty},\zeta) \in \Gr_{\HH_+}(\HH)$ is the closure of the space of functions $f\in L^2(S^1)$ admitting a meromorphic extension to $\bar{D}_0$ with pole divisor subordinate to $\DD$.

To realize the construction we use canonical theta-functions of $g$ variables, $g\geq 1$
\be \theta(\mathbf{z})=\sum_{\mathbf{m}\in \mathbb{Z}^g} \mathrm{exp}\left\{\imath \pi \mathbf{m}^T \Tb \mathbf{m} +2\imath \pi \mathbf{m}^T  \mathbf{z}
\right\},\quad \mathbf{z}\in\mathbb{C}^g,  \label{thetacan} \ee
where $\Tb$ is a complex symmetric $g\times g$-matrix with positive definite imaginary part. The space of such matrices (the Siegel upper half-space) will be denoted $\mathcal{S}^g$.  The theta-function is holomorphic on $\mathbb{C}^g\times \mathcal{S}^g$ and satisfies
\be
\theta(\mathbf{z}+\mathbf{n})=\theta(\mathbf{z}),\quad
\theta(\mathbf{z}+\Tb\mathbf{n})=\mathrm{exp}\{ -\imath\pi(\mathbf{n}^T\Tb\mathbf{n}+2\mathbf{z}^T\mathbf{n}) \}\theta(\mathbf{z})
.\label{thetatransf}\ee
 In the considered case of $\theta$-functions associated to an algebraic curve
 \be
 \Tb:=\grA^{-1}\grB,
 \ee
 where $\grA$ and $\grB$ are the period matrices of a basis of holomorphic differentials.

As shown in \cite{dub81, Kr77, djkm83}, the corresponding Baker-Akhiezer  function may be chosen
as the restriction to $\partial D_+$ of a meromorphic function on $X - p_\infty$ with
pole divisor $\DD$ and having an essential singularity at $p_\infty$ of the form
\be
\psi_w (p(\zeta),\tb) \sim e^{\sum_{i=1}^\infty t_i z^i}
 \left(1 + \OO\left({1\over z}\right)\right).
 \label{psilocalexp}
\ee

The Riemann-Roch theorem implies that there is just a one dimensional
space of such functions. They may be expressed, within a $\tb$-dependent normalization, as
\be
\tilde{\psi}_w(p, \tb)= e^{\int_{p_{{}_0}}^p \Omega(\tb)}
\frac{\theta (\AA(p) - \AA(\DD) +
 \sum_{i=1}^\infty \Ub_i t_i - \mathbf{K})}{
 \theta (\AA(p) - \AA(\DD) - \mathbf{K})},
 \label{tildepsiw}
\ee
where $\theta$ is the theta-function with $T$ equal to the Riemann matrix of periods
defined below relative to a
suitable polygonization obtained by cutting along a canonical homology basis
$(\mathfrak{a}_1, \dots , \mathfrak{a}_g, \mathfrak{b}_1, \dots , \mathfrak{b}_g)$ with intersection matrix
\be
\mathfrak{a}_i \circ \mathfrak{a}_j = \mathfrak{b}_i \circ \mathfrak{b}_j =0, \quad \mathfrak{a}_i \circ \mathfrak{b}_j = \delta_{ij},
\ee
$p_0$ is an arbitrarily chosen base point,
\bea
\AA:&& \SS^g(X) \ra \Cbb^g\nonumber \\
\AA:&=&\sum_{j=1}^gp_j\mapsto \sum_{j=1}^g\int_{p_0}^{p_j}\boldsymbol{\omega},
\quad \boldsymbol{\omega}=\left( \begin{array}{c}\omega_1\\ \vdots\\ \omega_g \end{array}\right)
 \label{abelmap}
\eea
is the Abel map with $i$th component
\be
\AA_i :=\sum_{j=1}^g \int_{p_{{}_0}}^{p_j} \omega_i,  \quad i=1, \dots , g,
\ee
where $(\omega_1, \dots , \omega_g)$ is a canonically normalized basis of the
space $H^0(K)$ of holomorphic abelian differentials
\be
\oint_{a_i }\omega_j = \delta_{ij}, \quad \oint_{b_i} \omega_j = T_{ij} ,
\ee
and $\mathbf{K} \in \Cb ^g$ is the Riemann constant, chosen so that $\theta (\AA(p) - \AA(\DD) -\mathbf{K})$
vanishes at the $g$ points $p= p_i$ in the divisor $\DD$.

Define the linear family of abelian differentials of the second type
\be
\Omega(\tb) = \sum_{j=1}^\infty \Omega_j t_j ,
\ee
where $\Omega_i$ is the unique normalized abelian differential  of the
second kind with pole divisor of degree $j+1$ at $p_\infty$ having local form
\be
\Omega_j\sim d(z^j)  + {\rm holomorphic}
\label{Omega_kdef}
\ee
near $p_\infty$, with vanishing $a-$cycles
\be
\oint_{\mathfrak{a}_i} \Omega_j = 0,  \quad i =1, \dots, g.
\label{Omega_knorm}
\ee
Then $2\pi i \Ub_j \in \Cbb^g$ is defined to be its vector of $b$-cycles with components
\be
\oint_{\mathfrak{b}_k} \Omega_j =: 2\pi i(\Ub_j)_k , \quad k=1, \dots, g.
\ee

In order to compare with the formal Baker function $\psi_a(z, \tb)$ appearing in the
Sato formula (\ref{sato}), we must interpret
$p=p(z)$ within the punctured disc $D_{\infty} - p_\infty$ and on its boundary, and normalize
$\tilde{\psi}(p, \tb)$ in formula (\ref{tildepsiw})
so as to obtain  the correct local expansion (\ref{psilocalexp}) near $\xi =0$
\be
\psi_w(p(\xi), \tb) = {\tilde{\psi}_w(p(\xi),\tb) \over a_0(\tb)},
\ee
where
\be
\tilde{\psi}_w(p(\xi), \tb) \sim e^{\int_{p_{{}_0}}^{p(\xi)}\Omega(\tb)}
 \left( a_0(\tb) + a_1(\tb)\xi+ \cdots \right).
\ee
Since $\int_{p_{{}_0}}^p \Omega_i$ has the local expansion
\be
 \int_{p_{{}_0}}^p \Omega_i = \xi^{-i} + \sum_{j=1}^\infty {1\over j}Q_{ij}\ \xi^j  + q_j  ,
 \label{intOmega_exp}
 \ee
 where
 \be
 Q_{ij } = Q_{ji},  \quad 1\le i,j \le \infty,
 \ee
 and $\AA(p(z))$ has the expansion \cite{dub81}, \cite{Dic03}
  \be
 \AA(p(z)) = \AA(p_\infty)  - \sum_{j=1}^\infty {1\over j} \Ub_j z^{-j},
 \label{abelexpansion}
 \ee
this gives the formula
\be
\psi_w(p(z), \tb)= e^{\sum_{i=1}^\infty t_i\left(z^i  + \sum_{j=1}^\infty {1 \over j }Q_{ji}z^{-j}\right)} \
{\theta(\eb+\sum_{i=1}^\infty \Ub_i(t_i - {1\over i}z^{-i})) \theta(\eb) \over  \theta(\eb-\sum_{i=1}^\infty  {1\over i}\Ub_iz^{-i})\theta(\eb +\sum_{i=1}^\infty \Ub_it_i )}
  \label{alggeomBaker}
\ee
near $z=\infty$ where
\be
\eb := \AA(p_\infty) - \AA(\DD)  - \mathbf{K}.
\label{abel_divisor_image}
\ee
(Note that the assumptions behind formula (\ref{tildepsiw}) imply that $\eb$ is not on the
theta divisor;  $\theta(\eb) \ne 0$.)
The last ratio of  theta function factors in (\ref{alggeomBaker}),

\[  \frac{\theta(\eb)}{ \theta \left( \eb+\sum_{i=1}^{\infty} \Ub_it_i \right)} \]
does not depend on
$z$, and hence the space spanned by the values of $\psi_w(p(z), \tb) $ is the same as that
spanned by
\bea
\check{\psi}_w(p(z), \tb) := e^{\sum_{i=1}^\infty t_i\left(z^i  + \sum_{j=1}^\infty {1 \over j }Q_{ji}z^{-j}\right)} \
{\theta(\eb+\sum_{i=1}^\infty \Ub_i(t_i - {1\over i}z^{-i}) )\over  \theta(\eb-\sum_{i=1}^\infty  {1\over i}\Ub_iz^{-i})}.
\label{checkalggeomBaker}
\eea
We may expand the remaining ratio of theta function terms as a power series in $z^{-1}$
\bea
\theta(\eb +\sum_{i=1}^\infty \Ub_i(t_i - \mbox{ {\small ${1\over i}$}}z^{-i})) &\&e^{\sum_{i=1}^\infty - {1\over i z^i} {\partial \over \partial t_i}}
\left(\theta(\eb +\sum_{i=1}^\infty \Ub_it_i )\right) \cr
&\& = \sum_{j=0}^\infty z^{-j} h_j(-\nabla_{\Ub}) \theta(\eb+\sum_{i=1}^\infty \Ub_it_i )
\eea
and
\be
\theta(\eb -\sum_{i=1}^\infty  \mbox{ {\small ${1\over i}$}}\Ub_iz^{-i})
 = \sum_{j=0}^\infty z^{-j} h_j(-\nabla_\Ub) \theta(\eb),
\ee
where
\be
\nabla_\Ub := (\nabla_{\Ub_1},  {1\over 2}\nabla_{\Ub_2},  {1\over 3}\nabla_{\Ub_3}, \dots)
\ee
and $\nabla_{\Ub_i}$ is the directional derivative in $\Cb^g$ along $U_i$.

Now define a basis $\{ w_0, w_1, \dots \}$ for $w$ as
\be
w_0 (z):=\check{\psi}(z, \tb)_{\tb=0} = 1, \quad
w_j(z) := \left({\partial \check{\psi}(z, \tb) \over \partial t_j}\right)_{\tb=0} , \quad j \ge 1.
\label{Baker_basis}
\ee
Then
\be
w_j(z) = z^j + P_{0j}+ \sum_{i=1}^\infty ( \mbox{ {\small ${1\over i}$}} Q_{ij} + P_{i j}) z^{-i},  \quad j=1, 2, \dots
\ee
where
\be
\sum_{i=0}^\infty P_{ij} z^{-i}  := {\sum_{i=0}^\infty M_{ij} z^{-i} \over \sum_{i=0}^\infty N_i z^{-i} }
\ee
and
\be
M_{ij} :=  \nabla_{\Ub_j} h_i(-\nabla_\Ub)\theta(\eb),  \quad
N_i:=  h_i(-\nabla_\Ub)\theta(\eb) .
\ee
It follows that the affine coordinates $A_{ij}$ of the element $w$ are
\bea
A_{i j} &\&:=\mbox{ {\small ${1\over i +1}$}} Q_{i+1, j} + P_{i+1,  j} ,  \quad i  = 0, 1 , \dots,
\quad  j=1, 2,  \dots  \cr
A_{i0} &\& =0, \quad i = 0, 1,2 \dots
\label{affine_quadratic}
\eea
By Corollary \ref{dethookplucker_coroll} this determines the Pl\"ucker coordinates
for all hook partitions and hence, by (\ref{dethookplucker}), for all partitions.

Comparing formula  (\ref{alggeomBaker}) for the Baker function with the  Sato formula
(\ref{sato}),  we see  that the tau function for $w=w(X,\mathcal{D},p_{\infty},\zeta)$ is given  by (cf. \cite{Dic03})
\be
\tau_w(\tb) = e^{\sum_{i=1}^\infty \lambda_ i t_i} e^{{-{1\over 2}} \sum_{i,j=1}^\infty Q_{ij} t_i t_j}
\theta( \mathbf{e} + \sum_{i=1}^\infty t_i \Ub_i),
\label{alg_curve_tau}
\ee
where
\be
\ \lambda_i   :=  \mu_i + i \sum_{k=0}^{i-1} {Q_{k, i-k} \over 2k (i-k)}\label{lambdai}
 \ee
with the $\mu_i$'s defined by
 \be
 \theta(\eb) e^{\sum_{i=1}^\infty {\mu_i \over i z^i}} := \theta(\eb- \sum_{i=1}^\infty {\Ub_i \over i z^i}) =\sum _{i=0}^\infty N_i z^{-i}.
\ee

  From the viewpoint of  the KP hierarchy, however, the linear exponential factor
  $e^{\sum_{i=1}^\infty\lambda_i t_i}$   in (\ref{alg_curve_tau}) may be removed,
  since this just corresponds to a  gauge transformation of the Baker-Akhiezer
  function  that is constant in the  $\tb$ variables
  \be
  \psi_w(z, \tb) \ra k(z)  \psi_w(z, \tb), \quad k(z) := e^{\sum_{i=1}^\infty{ \lambda_i \over i z^i}},
  \ee
  which leaves the solutions to the KP flow equations invariant.  Equivalently, this means
  replacing $w \in Gr_{\HH_+}(\HH)$ by
  \be
  w_k :=\mathrm{span} \{ k v, v\in w\}.
  \ee
   Therefore the tau function may be chosen in the simpler gauge equivalent form
  (cf. \cite{djkm83, fay83, Dic03})
\be
\tau_w(\tb) = e^{-{1\over 2} \sum_{i,j=1}^\infty Q_{ij}t_i t_j}
\theta( \eb + \sum_{i=1}^\infty t_i \Ub_i).
\label{alg_curve_tauQ}
\ee
Henceforth, we denote this $\tau$-function as $\tau(\eb, \tb)=\tau_{w}(\tb)$.

Applying  formula (\ref{schurderivbilin}) directly to $\tau_w(\tb)$ as defined
in (\ref{alg_curve_tauQ}) provides an alternative way to compute the affine coordinates
$A_{ab}$ that is equivalent, within such a gauge transformation, to (\ref{affine_quadratic}).

The Hirota bilinear relations (\ref{hirota1}) may in this case equivalently be written
\begin{eqnarray}
&&\mathrm{Res}_{\zeta=0} \frac{1}{\xi^2}
\left[\mathrm{exp}\left\{\sum_{n=1}^{\infty} t_n\xi^{-n} \right\}
\exp\left\{- \sum_{n=1}^{\infty} \frac{\xi^n}{n} \frac{\partial }{\partial t_n} \right\}\tau(\eb;\tb)
\right.\nonumber\\&& \left.
\qquad\qquad\quad
\times \mathrm{exp}\left\{-\sum_{n=1}^{\infty} \widetilde{t}_n\xi^{-n} \right\}
\exp\left\{\sum_{n=1}^{\infty} \frac{\xi^n}{n} \frac{\partial }{\partial \widetilde{t}_n} \right\}\tau(\eb;\widetilde{\tb})\right]=0,\label{bilinearresidue}
\end{eqnarray}
where $\xi=\xi(q)$ is the local coordinate of the point $q$ near $p$, $\xi(p)=0$ defining the KP hierarchy.

\subsection{Weierstrass gaps, bases, and the fundamental bi-differential}

The Weierstrass gap theorem  \cite{FK80} says:
\bt
[\bf L\"uckensatz]
For any point $p\in X$ on a nonsingular algebraic curve  $X$ of genus $g$ there exist precisely $g$ distinct non-negative integers $n_1,\ldots, n_g $, satisfying the inequalities
\begin{equation}
n_1=1<n_2<\ldots<n_g< 2g,
\label{wgaps}
\end{equation}
such that no meromorphic function on $X$ may have pole divisors solely at $p$  of degrees
$(n_1,\ldots,n_g)$.
\et

The set of integers $(n_1, \dots , n_g)$ is called the Weierstrass gap numbers,  and will
be denoted $\mathfrak{W}(p)$.
For a point in ``general position'' the gap numbers are $1,2,\ldots,g$.
A point $p\in X$ that admits a meromorphic function that has a pole of order smaller then $g+1$ is called a Weierstrass point.

Given the point $p_\infty$,  and local parameter $\xi(p)$,  the $2g$ dimensional space
$\mathrm{H}^*_1( X,\mathbb{Z} )$ dual  to the homology group  $\mathrm{H}_1( X,\mathbb{Z} )$ may be identified with a space of meromorphic differentials having poles at $p_\infty$ only, with vanishing residues, the pairing being given by integration over cycles.
A basis $\{u_1,\ldots,u_g, \Omega_{n_1}, \dots \Omega_{n_g}\}$ for this space consists of
the $g$ elements  $\{u_1,\ldots,u_g\}$ providing
a basis for  the subspace  $\mathrm{H}^0(X,K)$  of holomorphic  differentials,
defined so that  $u_j$  vanishes to order $n_j-1$ at $p_\infty$
\be
u_k=-\left(\xi(p)^{n_k-1} + \rm{higher}\;\rm{ order}\;\rm{ terms} \right) \mathrm{d}\xi(p), \quad k=1,\ldots,g,\quad n_k\in \mathfrak{W}(p),
\label{uholomorphicexp}
\ee
with higher order terms $\xi(p)^k$ consisting only of powers $k$ for which $k+1 \notin
 \mathfrak{W}(p)$.

The remaining basis elements $\{\Omega_{n_1},\ldots,\Omega_{n_g}$
are the  normalized differentials of second kind
 where the $\Omega_k$'s are as defined in (\ref{Omega_kdef}),  (\ref{Omega_knorm}).
 These are mutually dual under the pairing
    \be
\frac{1}{2\pi i}{\int \int}_X  u_j \wedge \Omega_{n_k} = \res_{p=p_\infty}\left( \left(\int_{p_0}^p u_j \right) \Omega_{n_k} (p)\right) = \delta_{jk}.
  \label{respairing}
\ee

Denote the non-vanishing period matrices around the $\mathfrak{a}$ and $\mathfrak{b}$ cycles
\bea
\oint_{\mathfrak{a}_j} u_i &\&=: \mathfrak{A}_{ij},\quad \oint_{\mathfrak{b}_j} u_i =: \mathfrak{B}_{ij},  \label{abper}\\
\quad {1\over  2\pi i}  \oint_{\mathfrak{b}_j} \Omega_{n_i} &\&=: C_{ij} = (U_{n_i})_j,
  \qquad i,j=1,\ldots,g.
  \label{bilinearrelations}
\eea
The columns of  the $g \times g$ matrix $\Cb$ are thus given by the vectors $\Ub_{n_i}$ corresponding to the Weierstrass gaps
\be
\Cb=(\Ub_{n_1},\Ub_{n_2},\ldots, \Ub_{n_g}),\quad n_j\in\mathfrak{W}(p_\infty).
   \ee

The Riemann bilinear relations, obtained by applying Stokes theorem  to the 2-forms $u_j \wedge u_k$  and  $u_j \wedge \Omega_k$  on the canonical polygonization of the curve, then imply
\bea
\grA \grB^T&\&= \grB  \grA^T, \cr
\grA \Cb^T &\& =  \boldsymbol{1}_g.
   \label{Riemann_bilin}
\eea
The relation to the basis $\{\omega_1, \dots \omega_1, \Omega_{n_1}, \dots , \Omega_{n_g}\}$
of normalized differentials  is thus
\be
\omega_i = \sum_{j=1}^g  \grA^{-1}_{ij} u_j = \sum_{j=1}^g C_{ji} u_j,
    \ee
and the normalized Riemann period matrix $\Tb$ is
 \be
 \Tb:= \grA^{-1}\grB= \Cb^T\grB.
 \ee
 Defining the vectors
 \be
 \Rb_ j= \grA \Ub_j, \quad j=1, 2, \dots,
 \ee
it follows from ({\ref{abelexpansion}) that the differentials $u_i$ have the local expansion
\be
u_i(p(\xi)) = -\sum_{j=n_i}^\infty (\Rb_j)_i \xi(p)^{j-1} d\xi(p).
\ee

It is also convenient to introduce the normalized symmetric bi-differential
 $\Omega(p,q)$ on $X \times X $, \cite{kl86},\cite{kl88}, \cite{fay73}
 defined by the conditions
\begin{itemize}
\item $\Omega(p,q)$ has a second order pole on the diagonal $p=q$ where its
local form, expressed in terms of the parameters $\xi(p),  \xi(q)$ is
\be
    {\Omega}(p,q)=\left( \frac{1}{(\xi(q)-\xi(p)^2} +\sum_{i,j=0}^{\infty} \mu_{ij}\xi(p)^i \xi(q)^j  \right)\mathrm{d}\xi(p)\mathrm{d}\xi(q)
    \label{bidifferentialpole}
\ee
    with
    \be
 \mu_{ij} =\mu_{ji}.
    \ee
    \item $\Omega(p,q)$ is holomorphic elsewhere in both $p$ and $q$.
        \item The $a$-cycles all vanish
    \be \
    \oint_{p\in {\mathfrak{a}_j}}\Omega(p,q) =  \oint_{q\in {\mathfrak{a}_j}}\Omega(p,q) = 0.\label{omegaavanish}
    \ee
\end{itemize}

  These conditions uniquely determine $\Omega(q,p)$,  which may  be given the
  explicit representation
  \be
  \Omega(p,q) = \mathrm{d}_p \mathrm{d}_q \mathrm{ln}\,\theta( \AA(p) - \AA(q)+\boldsymbol{\delta}),
  \ee
 where $\boldsymbol{\delta}$ is non-singular odd half-period.

The following further properties also follow from the definitions.
  \begin{itemize}
  \item The $b$-cycles are given by the normalized holomorphic differentials $\omega_j$, $j=1, \dots g$
  \be
  \oint_{p\in \mathfrak{b}_j} \Omega(p,q) =2\pi i \omega_j(q), \quad \oint_{q\in \mathfrak{b}_j} \Omega(p,q)= 2\pi i  \omega_j(p).
  \label{Omega2bcycle}
  \ee
    \item The residues of the locally defined  bi-differentials $\xi(p)^{-j}\Omega(p,q)$,
    $\xi(q)^{-j}\Omega(p,q)$
    at $ p_\infty$ are given by the normalized second type differentials $\Omega_j$,
    \be
    \res_{p=p_\infty} \xi(p)^{-j}\Omega(p,q) = -\Omega_j(q),
    \quad     \res_{q=p_\infty} \xi(q)^{-j}\Omega(p,q) = -\Omega_j(p).
    \ee
    \label{Omegares}
  \item The coefficients $\mu_{ij}$ in the expansion (\ref{bidifferentialpole}) are related those
in  the expansion (\ref{intOmega_exp}) as follows
  \be
\mu_{ij} = - Q_{i+1, j+1}.
  \label{muQident}
  \ee
\end{itemize}

\subsection{Planar model of the curve}

 Henceforth, we assume the algebraic curve $X$ of geometric genus $g\geq 1$ is given by the equation
\begin{equation}
X: \; P(x,y)=0,\quad  x,y\in\Cb  ,
\end{equation}
with $P$ a polynomial in $x$ and $y$
\begin{equation}
P(x,y)=y^n+a_{1}(x)y^{n-1}+\ldots+a_n(x),
\label{planarcurve}
\end{equation}
where $a_k(x),k=0\ldots,n$ are polynomials in $x$ and $n>1$.
Suppose that the curve $X$ has a Weierstrass point at infinity, $p_{\infty}$ where coordinates $x,y$ are locally expressed as
\be x=\frac{1}{\xi^n}+\ldots,\quad y=\frac{1}{\xi^s}+\ldots  \label{local}\ee
and the order of any monomial term  the polynomial $P(x,y)$ is the order of its pole at $p_{\infty}$. Now suppose that the curve $X$ can be written in the form
\be   y^n-x^s+\quad \rm{lower}\;\;\rm{ order}\;\; \rm{terms} \; =0,   \label{ns}  \ee
where $n,s >2$ are positive integers .

In what follows the planar coordinates of a point $p$  are denoted $(x(p),y(p))$.
 When considering local expansions near to a reference point $p_\infty$ with local parameter $\xi(p)$
 we also use $p(\xi)$ to denote the point, with $p(0)= p_\infty$. It is then possible to include the principal part of $\Omega$ in an explicit algebraic expression in terms of the coefficients of the curve $X$.
\bt[ \cite{ba97}]
\label{fundamental_diff}
The fundamental bi-differential may be expressed in the form
\bea
\Omega(p,q)&\&= {\mathcal{F}(p,q)\over (x(p) - x(q))^2} {\mathrm{d} x(p)  \mathrm{d} x(q) \over P_y(x(p), y(p)) P_y(x(q),y(q))} + \sum_{i=1}^g\sum_{j=1}^g \omega_i(p)\gamma_{ij}\omega_j(q) \label{fundamentalOmega0} \\
&\& = {\mathcal{F}(p,q)\over (x(p) - x(q))^2} {\mathrm{d} x(p)  \mathrm{d} x(q) \over P_y(x(p), y(p)) P_y(x(q),y(q))} + \sum_{i=1}^g\sum_{j=1}^g u_i(p)\varkappa_{ij}u_j(q)\label{fundamentalOmega}
\eea
where $\mathcal{F}(p,q)$ is a polynomial function of the coordinates  $(x(p), y(p), x(q), y(q))$ which is
a linear form in the coefficients of the polynomials $\{a_0(x) , \dots a_n(x)\}$ defining the planar model of the curve (\ref{planarcurve}), and the symmetric $g \times g$ matrices $\varkappa$,  $\gamma$ have elements $\varkappa_{ij}$,  $\gamma_{ij}$  given by
\be
\gamma_{ij}=(\grA^T \varkappa \grA)_{ij} = -\oint_{p\in \mathfrak{a}_i} \oint_{q\in \mathfrak{a}_j}{\mathcal{F}(p,q)\over (x(p) - x(q))^2} {\mathrm{d} x(p)  \mathrm{d} x(q) \over P_y(x(p), y(p)) P_y(x(q),y(q))} .
\label{kappadef}
\ee
\et

\br The representation (\ref{fundamentalOmega}) of the fundamental bi-differential $\Omega(p,q)$ in the hyperelliptic case is classical and may be found  in the books \cite{ba97} and \cite{ba07}. It is summarized  in \cite{bel97b} and extended to non-hyperelliptic curves in \cite{eel00}, \cite{bel00}.
\er

\br As already mentioned above in Remark 1.1,  the matrix $\boldsymbol{\varkappa}$ appearing in the definition of the multi-variable $\sigma$-function is defined only up to the addition of an arbitrary symmetric matrix, say $\boldsymbol{\chi}$. The change $\boldsymbol{\varkappa} \rightarrow \boldsymbol{\varkappa}+ \boldsymbol{\chi}$, however, does not affect the higher Klein formula nor  the algebraic and differential relations between the $\wp$-functions that follow from this formula. In the  examples to follow, explicit expressions will be given for the polynomial $\mathcal{F}(p,q)$ defining the fundamental bi-differential $\Omega(p,q)$.
\er

The following is a consequence of  (\ref{fundamentalOmega}).

\bc
\label{cor31}
The coefficients $\mu_{ij}$ in the expansion  (\ref{bidifferentialpole}) can be decomposed as a sum
\be
\mu_{ij}= - Q_{i+1, j+1}=\mu_{ij}^{\rm{alg}}+\mu_{ij}^{\rm{trans}},\label{summu}
\ee
where $\mu_{ij}^{\rm{alg}}$ is defined by the expansion
\begin{eqnarray}
 \Omega^{\rm{alg}}(p,q)&=&{\mathcal{F}(p,q)\over (x(p) - x(q))^2} {\mathrm{d} x(p)  \mathrm{d} x(q) \over P_y(x(p), y(p)) P_y(x(q),y(q))}\nonumber \\
 \nonumber\\
 &=&\left( \frac{1}{(\xi(q)-\xi(p))^2} +\sum_{i,j=0}^{\infty}\mu^{\rm{alg}}_{ij}\xi(p)^i \xi(q)^j  \right)\mathrm{d}\xi(q)\mathrm{d}\xi(p),
    \label{algbidifferentialexp}
\end{eqnarray}
$\xi(p)$ and $\xi(q)$ are local coordinates of the points $p,q$ and $\mu_{ij}^{\rm{trans}}$ is given by
\be
\mu_{ij}^{\rm{trans}}= \Rb^T_{i+1}\varkappa \Rb_{j+1}.\label{mutrans}
\ee
\ec
{\bf Proof:} The coefficients $\mu_{ij}$ in the expansion of the normalized symmetric bi-differential $\Omega(p,q)$ given in (\ref{bidifferentialpole})  (projective connection)  near the diagonal $p=q$ can be expressed as a sum $\mu_{ij}=\mu_{ij}^{\rm{alg}}+\mu_{ij}^{\rm{trans}}$. The first term is obtained by expansion of the l.h.s of (\ref{algbidifferentialexp}) as a rational bi-form on $X\times X$. The second term follows from the local expansion in powers of $\xi$ of the holomorphic differentials $u_i$ in the second term of the r.h.s of (\ref{fundamentalOmega}). It is transcendental and given by (\ref{mutrans}).
The term $\mu_{ij}^{\rm{alg}}$ defines the holomorphic part of the expansion of the rational 2-form appearing in the first term in (\ref{fundamentalOmega0}), (\ref{fundamentalOmega}).

\br Note that the double integral in (\ref{kappadef}) can be decomposed
into periods of basic holomorphic and meromorphic second kind differentials. Following  the Baker construction \cite{ba97},\cite{ba98}, we  represent the  integrand
$\Omega^{\rm{alg}}(p,q)$ in the form
\bea
\Omega^{\rm{alg}}(p,q)=\frac{\mathcal{F}((x,y),(z,w))}{(x-z)^2}
\frac{\mathrm{d}x\mathrm{d}z}{P_y(x,y)P_z(z,w)}\nonumber\\
=\frac{\partial}{\partial z} \Pi_{(z,w)}^{(z',w')}(x,y) dz
+\mathbf{u}^T(x,y)\mathbf{r}(z,w).\label{thirdkind}
\eea
Here $\Pi_{(z,w)}^{(z',w')}(x,y)$ is a third kind differential with first order poles in points $(z,w)$ and $(z',w')$ and corresponding residues $\pm 1$, $\mathbf{u}=(u_1,\ldots,u_g)^T$ is the vector of  holomorphic differentials normalized as in eq.~(\ref{uholomorphicexp}),    $\rb=(r_1,\ldots,r_g)^T$ is the vector of meromorphic differentials with poles of orders $n_1+1,\ldots,n_g+1$ at $p_{\infty}$.  (The pole location $(z',w')$ can be taken arbitrary and does not affect the construction.) The differentials $\rb$ are then chosen to satisfy the symmetry condition
\be  \Omega^{\rm{alg}}(p,q)=\Omega^{\rm{alg}}(q,p). \ee
The explicit algebraic construction of the differentials $\mathbf{r}$ is described in \cite{ba97} and further developed in \cite{bel97b}. In particular, in the case of a hyperelliptic curve
\be  P(x,y)=y^2-\mathcal{P}_{2g+1}(x), \ee
where $\mathcal{P}_{2g+1}(x)$ is a polynomial in $x$ of degree $2g+1$,
\be
 \Pi_{(z,w)}^{(z',w')}(x,y)=\frac{1}{2y} \left\{  \frac{y+w}{x-z}  - \frac{y+w'}{x-z'}  \right\}\mathrm{d}x \label{hypthirdkind}
\ee
The second term in (\ref{hypthirdkind})  can be taken as an arbitrary finite point of the curve.
\er
The matrices of $\mathfrak{a}$ and $\mathfrak{b}$-periods of the differentials $\rb$ as  given in the introduction by eqs.~(\ref{stmat}) and (\ref{varkappa}) enter in  the definition of the multi-variable 
$\sigma$-function (\ref{sigmadef}) below.


\subsection{$\sigma$-functions and algebro-geometric formulae for $\pi_{\lambda}(w)$}
Let $\mathcal{D}=p_1+\ldots+p_g$ be a positive non-special divisor of degree $g$ and
\be
 \mathbf{v}:=\sum_{i=1}^g\int_{p_{\infty}}^{q_i} \mathbf{u}+{\grA}\mathbf{K}=-\grA\eb,
 \label{abelmap}
 \ee
where $\grA$ is defined in (\ref{bilinearrelations}), $\mathbf{K}$ is the vector of Riemann constants, with the base point at $p_{\infty}$ and $\mathbf{u} =(u_1,\ldots, u_g)^T$.

The multivariable
$\sigma$-function $\sigma(\vb)$  is defined by the formula
\be
\sigma(\mathbf{v})=C\theta(\eb )\mathrm{exp}\left\{ \frac12 \mathbf{v}^T \varkappa \mathbf{v}  \right\},\quad \eb=-\grA^{-1} \vb
\label{sigmadef}
\ee
and $C$ is a constant depending on the moduli of the curve,  whose explicit form is not needed here. This definition naturally generalizes the Weierstrass $\sigma$-function from the theory of elliptic functions to higher genera.

The multi-variable $\boldsymbol{\zeta}$-functions, $\boldsymbol{\zeta}=(\zeta_1,\ldots, \zeta_g)$ are defined as
\be  \zeta_{k}\left(\mathbf{v}\right)=\frac{\partial}{ \partial v_k}\mathrm{ln}\;\sigma(\mathbf{v}), \quad k=1,\ldots,g . \label{kleinzeta} \ee
The Kleinian $\wp$-functions are the second logarithmic derivatives of the sigma function
\begin{eqnarray}
\wp_{i,k}\left(\mathbf{v}\right)=-\frac{\partial^2}{ \partial v_i \partial v_k}\mathrm{ln}\;\sigma(\mathbf{v}), \quad i,k=1,\ldots,g \label{kleinwp}
\end{eqnarray}
More generally, we denote higher order logarithmic derivatives as
$$\wp_{\underbrace{i_1,\ldots,i_1}_{m_1},\ldots,\underbrace{i_k,\ldots,i_k}_{m_k}}(\vb) =- \frac{\partial^{m_1i_1+\ldots+m_ki_k}}{\partial
v_{i_1}^{m_1}\cdots \partial v_{i_k}^{m_k} }\mathrm{ln}\,\sigma (\vb),\quad i_1,\ldots,i_k\in \{1,\ldots,g\}.$$

In the classical theory the following theorem provides the basic means to derive algebraic and differential relations between multi-variable $\wp$-functions. (See e.g. \cite{ba97} and the more recent exposition \cite{bel97b}).

\begin{theorem}[\bf Klein formula]  Let the planar curve $X$ of genus $g$ be defined by the polynomial equation $P(x,y)=0$. Choose a set of independent holomorphic differentials in the form
\begin{equation}
u_k=\frac{\phi_k(x,y)}{f_y(x,y)}dx,\quad k=1,\ldots, g,
\end{equation}
where $\phi_k(x,y)$ are polynomials in $x$ and $y$. Let $p=(x,y)$ be an arbitrary point of $X$ and $\mathcal{D}=p_1+p_2+\ldots+p_g$
a positive non-special divisor on $X$, $p_k=(x_k,y_k)$ and $p=(x,y)\in X$ - arbitrary point. Let  $\mathbf{v}$ be
the shifted image of $\mathcal{D}$ under the Abel map as in (\ref{abelmap}). Then
\begin{eqnarray}
\sum_{j,k=1}^g  \wp_{j,k}\left( \int_{p_{0}}^p \mathbf{u}-\mathbf{v}\right)\phi_k(x,y)\phi_j(x_r,y_r)=\frac{\mathcal{F}(p,p_r)}{(x-x_r)^2},\quad r=1,\ldots,g, \label{kleinformula}
\end{eqnarray}
where the polynomial $\mathcal{F}(p,p_r)=\mathcal{F}((x,y);(x_r,y_r))$
defines the fundamental bi-differential $\Omega(p,p_r)$.
\end{theorem}

\br
This formula was first given for hyperelliptic curves in \cite{kl86}, \cite{kl88}. But the proof, which is 
based on the Riemann vanishing theorem and the representation of the fundamental bi-differential in the form (\ref{fundamentalOmega}) can easily be extended to non-hyperelliptic curves.

The Weierstrass-Poincar\'e theorem  (see e.g.~\cite{ig82}) says that any $g+1$ Abelian functions on the Jacobian of an algebraic curve of genus $g$ are algebraically dependent. In particular, it follows from this that in the case of a hyperelliptic curve of genus $g$, among the $(g+1)g/2$ functions $\wp_{ij}$ there are only $g$ that are algebraically independent,  so  these functions must  satisfy  $g(g-1)/2$ relations. (Using the Klein formula (\ref{kleinformula}) it was shown in  [BEL97] that these relations are quartics that represent a Kummer variety. (See the example of a genus $2$ curve in Section 4.1 below).)

Another set of relations that follow from the Klein formula describe the Jacobi variety of the curve $X$ and integrable flows of KP type using $\wp$-functions as coordinates. In particular in the case of hyperelliptic curves all products $\wp_{ijk}\wp_{p,q,r}$ are cubic polynomials in $\wp_{ij}$. (or more details see [BEL97].)
Here we will show that these results can also be obtained within $\tau$-function theory on the basis of the Sato formula or, equivalently from the bilinear Identity. To do so we represent the Sato $\tau$-function in terms of the multi-variable $\sigma$-function of Klein. These methods of derivation of integrable hierarchies of KP type, Jacobi and Kummer varieties are compared  in \cite{eeg10}.
\er
\bp
\label{tau_sigma_prop}
The normalized algebro-geometric function $\tau(\eb, \tb;)/\tau(\eb, \mathbf{0};)$  is expressible
 in terms of the multi-variable $\sigma$-function as
\begin{eqnarray}
\frac{\tau(\eb, \mathbf{t})}{\tau(\eb, \mathbf{0})}=\frac{\sigma\left( \sum_{k=1}^{\infty} \mathbf{R}_k t_k +\vb \right)}{\sigma(\vb)}
 \mathrm{exp}\left\{ \sum_{k=1}^{\infty}\Lambda_k(\vb) t_k  \right\}
 \mathrm{exp}\left\{ \frac12 \sum_{k,l=1}^{\infty}\mu_{kl}^{\rm{alg}} t_kt_l \right\},
 \label{tau_sigmae1}
\end{eqnarray}
Here $\vb=\grA\eb$ is the shifted Abelian image (\ref{abelmap}) of the positive non-special divisor $\mathcal{D}$ and $\mu_{ik}^{\rm{alg}}$ are coefficients in the expansion of the algebraic part of the bi-differential $\Omega(p,q)$ near the point $p_{\infty}$. The coefficients $ \Lambda_k(\vb)$ are given by
\be
 \Lambda_k(\vb)=\mathbf{R}^T_k\varkappa \vb  \,\quad k=1,2,\ldots\label{lambdabig}\ee.
\ep
{\bf Proof}. The algebro-geometric $\tau$-function in the gauge transformed form (\ref{alg_curve_tauQ}) leads to the relation
\bea
\frac{\tau(\eb, \mathbf{t})}{ \tau(\eb, \mathbf{0}) }= \mathrm{exp}\left\{\frac12 \sum_{i,j=0}^{\infty}\mu_{ij}^{\rm{alg}}t_it_j+ \frac12\sum_{i,j=0}^{\infty}\mu_{ij}^{\rm{trans}}t_it_j \right\}\frac{\theta(\eb+\sum_{i=1}^{\infty}\mathbf{U}_it_i)}{ \theta(\eb)  }
\eea
where $\lambda_i$ is given in (\ref{lambdai}) and the relation (\ref{muQident}) was used.

Otherwise from the definition of $\sigma$-function we get
\begin{eqnarray}
\sigma\left( \sum_{k=1}^{\infty} \mathbf{R}_kt_k+\vb \right)
&=& C \theta\left( \sum_{k=1}^{\infty} \Ub_kt_k+\grA^{-1}\vb \right)\nonumber\\
&\times&\mathrm{exp}\left\{ \frac12 \sum_{k,l=1}^{\infty} \mathbf{R}_k^T \varkappa \mathbf{R}_l t_kt_l  \right\} \label{sigmatauexp}\\
&\times& \mathrm{exp}\left\{ \sum_{k=1}^{\infty} \mathbf{R}_k^T\varkappa \vb\, t_k  \right\}\mathrm{exp}\left\{ \frac12 \vb^T\varkappa \vb\right\}.\nonumber
\end{eqnarray}
Here  $C$ is the constant given in the definition of the sigma-function (\ref{sigmadef}).
Taking into account  eq.~(\ref{mutrans}) we get (\ref{tau_sigmae1}) with the coefficients $\Lambda_k$ given in  (\ref{lambdabig}).

Since the $\tau$-function is defined only up to a linear exponential factor in $\mathbf{t}$, we omit the linear term in the exponential (\ref{tau_sigmae1}), to get the simpler formula
\begin{eqnarray}
\frac{\tau(\eb, \mathbf{t})}{\tau(\eb, \mathbf{0})}=\frac{\sigma\left( \sum_{k=1}^{\infty} \mathbf{R}_k t_k +\vb \right)}{\sigma(\vb)}
 \mathrm{exp}\left\{ \frac12 \sum_{k,l=1}^{\infty}\mu_{k,l}^{\rm{alg}} t_kt_l \right\}, \quad \vb=\grA\eb.\label{tau_sigmae}
\end{eqnarray}

\br A similar formula for the algebro-geometric $\tau$-functions in terms of the $\sigma$-function was given by A.Nakayashiki in \cite{nak09}, where the terms linear in $t_k$ in the exponent are also taken into account.
\er

 At first glance the use of the multi-variable sigma-function instead of the Riemann theta-function in the expression for the algebro-geometric $\tau$-function seems a trivial change. But as a result, the quadratic form in the
exponential has coefficients that are algebraically expressed in terms of coefficients of the curve. Moreover, these coefficients, $\mu_{kl}^{\rm{alg}}$ are polynomials in the coefficients of the curve (see \cite{nak08} for details).  As shown below, such a representation of the $\tau$-function results in solutions of the integrable KP hierarchy expressed as differential polynomials in the $\wp$-functions with polynomial coefficients determined directly in terms of the coefficients of the polynomials $P(x,y)$ defining the curve.

According to Propositions \ref{pluckercoordinate} and \ref{satoprop}
for any curve $X$ of genus $g$, the associated algebro-geometric $\tau$-function admits the following expansion
\be
\frac{\tau(\eb, \boldsymbol{t})}
  {\tau(\mathbf{\eb, 0})} = \sum_{\lambda}\pi_{\lambda}(w)s_{\lambda}(\mathbf{t}) ,\label{tauexp} \ee
where
\be
\pi_{\lambda}(w)= (-1)^{\sum_{k=1}^r b_k} \det(A_{a_i,b_j}|_{\leq i,j\leq r}),
\ee
the sum being over all partitions $\lambda$, which in Frobenius notations have the form $\qquad$  $(a_1,\ldots, a_r |  b_1, \ldots,  b_r)$. Here $A_{ij}$ with $i,j=0\ldots, \infty$ are the elements of the affine coordinate matrix $A$ representing the Grassmannian element $w(X,\mathcal{D},p_{\infty},\zeta)\in  \mathrm{Gr}_{\HH_+}(\HH)$.

\bc
The elements $A_{ij}$ are expressible as polynomials in the Kleinian symbols
\begin{equation}  \zeta_i(\vb),\; \;\wp_{ij}(\vb),\;\ldots,\quad i,j \in\{1,\ldots,g\} \label{coordinates}\end{equation}
and the coefficients of the polynomial $P(x,y)$.
\ec

\noindent{\bf Proof}. The quantities $\pi_{\lambda}(w)$   in the $\tau$-expansion (\ref{tauexp}) are expressible in terms of quotients $\sigma_i(\vb)/\sigma(\vb)$, $\sigma_{ij}(\vb)/\sigma(\vb)$, etc. But for
$i,j,k\ldots=1,\ldots g$ we get
\begin{eqnarray}
\frac{\sigma_i(\vb)}{\sigma(\vb)}&=&\zeta_i(\vb),\quad \frac{\sigma_{ij}(\vb)}{\sigma(\vb)}=\zeta_i(\vb)\zeta_j(\vb)-\wp_{ij}(\vb),\\
\frac{\sigma_{ijk}(\vb)}{\sigma(\vb)}
&=&\zeta_i(\vb)(\vb)\wp_{j,k}(\vb) +\zeta_j(\vb)(\vb)\wp_{i,k}(\vb)+\zeta_k(\vb)(\vb)\wp_{ij}(\vb)\cr
  &&-\zeta_i(\vb)\zeta_j(\vb)\zeta_k(\vb)+\wp_{ijk}(\vb).\\
&\vdots & \nonumber
\end{eqnarray}

To each symbol $\wp_{k_1,\ldots,k_g}$  we assign a weight
\be
\wp_{\underbrace{\scriptstyle 1,\ldots, 1}_{k_1},
  \underbrace{\scriptstyle 2,\ldots, 2}_{k_2}, \ldots,
  \underbrace{\scriptstyle g,\ldots, g}_{k_g}}\quad \Leftrightarrow
\quad \mathcal{W}_{k_1\ldots k_g}= \sum_{j=1}^gk_jn_j,
\label{weightP}\ee
where $\{n_i\}_{i=1,\ldots,g}$ is the Weierstrass gap sequence at infinity.

To each coefficient $a_{kl}$ of a monomial term  $a_{kl}x^ky^l$, $k<s, l<n $ within the polynomial $P(x,y)$ defining the curve  (\ref{planarcurve}) we assign the weight $\widehat{\mathcal{W}}_{kl}$,
\be   a_{kl} \quad \Leftrightarrow
\quad  \widehat{ \mathcal{W}}_{kl}=ns-(nk+ls).   \label{weightA} \ee

Finally,  assign to a monomial whose factors are $\wp$-symbols and coefficients $a_{k,j}$ the weight $\mathcal{W}$ that is the sum of weights of factors,
\be
a_{ij}\cdots a_{kl} \wp_{i_1\ldots i_g}\cdots \wp_{k_1\ldots k_g} \quad \Leftrightarrow \quad \mathcal{W}=\widehat{\mathcal{W}}_{ij} +\cdots +\widehat{\mathcal{W}}_{kl}+\mathcal{W}_{i_1\ldots i_g}+\cdots+\mathcal{W}_{k_1\ldots k_g}.
\ee
Consider the set of (Giambelli-like) relations
\be
\pi_{(a_1,\ldots,a_r| b_1,\ldots,b_r)}=(-1)^{\sum_{i=1}^r b_i}\mathrm{det}(A_{a_i,b_j})\label{Giambelli1}
\ee
corresponding to a partition $\lambda=(a_1,\ldots,a_r| b_1,\ldots,b_r)$ of weight $\mathcal{W}$. The above procedure reduce relations (\ref{Giambelli1})  to corresponding homogeneous polynomial relations of weight $\mathcal{W}$ between $\wp$-functions, $\wp_{i_1,\ldots, i_k}$ with coefficients that are polynomials in the coefficients of the defining curve polynomial $P(x,y)$. Such relations describe KP-type hierarchies in terms of $\wp$-coordinate.


\section{Examples and applications of Schur function expansions}


\subsection{$\tau$-function of a hyperelliptic curve}
Let $X$ be a hyperelliptic curve of genus $g$ defined by the equation
\be X: \quad P(x,y)=0 \ee
with polynomial $P(x,y)$ given as
\bea
P(x,y)&=&y^2-4x^{2g+1}
+\ldots+\alpha_0\cr
&=&y^2-4\prod_{j=1}^{2g+1}(x-a_j)
\label{hyperel}.
\eea
As above,  denote by $p=(x,y)$ an arbitrary point of $X$ and $p_{\infty}=(\infty,\infty)$. We choose a canonical basis of cycles
$(\gra_1,\ldots,\gra_g; \grb_1,
\ldots, \grb_g)\in H_1(X,\mathbb{Z})$
and fix the basic holomorphic differentials
$\mathbf{u}=(u_1,\ldots,u_g)^T$ as
\begin{equation}
u_i(p)=\frac{x^{i-1}\mathrm{d} x}{y}, \quad i=1,\ldots,g.
\end{equation}
Denote as above period matrices, $\grA$, $\grB$ and $\Tb=\grA^{-1}\grB$.

Let $\mathcal{D}=p_1+\ldots+p_g$ be non-special divisor of degree $g$ and let $\vb$
be the shifted Abel map given in (\ref{abelmap}). In this case the vector of Riemann constants $\mathbf{K}$ can be given as the image under the Abel map  of the divisor $\mathcal{D}_{\kappa}=(p_1,\ldots,p_g  )$ where $p_k=(a_k,0)$ are branch points whose abelian images are odd half-periods [FK80, Sect VII.1.2],
\be
 \mathbf{K} = -\sum_{ j=1}^{g} \int_{p_0}^{p_k}\boldsymbol{\omega}= -\grA^{-1}\sum_{ j=1}^{g} \int_{p_0}^{p_k} \mathbf{u}
 \ee
Therefore we can write
\be
\mathbf{v}= \sum_{k=1}^g   \int_{ (a_k,0)}^{q_k} \mathbf{u}.  \label{vhyperel}
 \ee

In the classical theory ( see e.g. \cite{ba97,ba98}) it was shown that the quadratic bi-differential $\Omega(p,q)$ can be chosen in the form
\begin{eqnarray}
\Omega(p,q)=\frac{F(x,z)+2yw}{4(x-z)^2yw} \mathrm{d} x \mathrm{d} z +2\mathbf{v}^T(p)\boldsymbol{\varkappa} \mathbf{v}(q), \label{hyperelomega}
\end{eqnarray}
where the polynomial $F(x,z)$ is the Kleinian 2-polar
\begin{equation}F(x,z)=\sum_{m=0}^g x^mz^m ( 2\alpha_{2m}+(x+z)\alpha_{2m+1} )  \label{2polar}\end{equation}
and the $g\times g$-matrix symmetric matrix $\varkappa$ is given as $\boldsymbol{\varkappa}=\grA^{-1}\grS$, where $\grS$ the matrix of $\mathfrak{a}$-periods, of meromorphic differentials
\begin{equation}
r_j=\sum_{k=j}^{2g+1-j} (k+1-j)
   \lambda_{k+1+j} \frac{x^k \mathrm{ d}x }{ 4y},   \quad j=1,
  \ldots, g.\label{hypmer}
\end{equation}

\begin{theorem}[\bf Klein formula for hyperelliptic curve]  Let the planar curve $X$ be defined by the polynomial equation (\ref{hyperel}). Let $p=(x,y)$ be an arbitrary point of $X$ and $\mathcal{D}=p_1+p_2+\ldots+p_g$
a non-special divisor on $X$, $p_k=(x_k,y_k)$. Let  $\mathbf{v}$ be the shifted
Abel map of $\mathcal{D}$ given in (\ref{vhyperel}). Then
\begin{eqnarray}
\sum_{i,k=1}^g  \wp_{i,k}\left( \mathcal{A}(p)-\mathcal{A}(p_{\infty})+\mathbf{v}\right)x^{k-1}x_r^{i-1}=\frac{F(x,x_r)-2yy_r }{4(x-x_r)^2},\quad r=1,\ldots,g,
\end{eqnarray}
where the polynomial $\mathcal{F}(p,p_r)=\mathcal{F}((x,y);(x_r,y_r))$
defines the fundamental bi-differential $\Omega(p,p_r)$.
\end{theorem}

\br
Differential relations between the $\wp_{ij}$ describe all possible integrable equations associated with the given curve.
Moreover one can show that for arbitrary genus $g$ any even derivative
$\wp_{i_1,\ldots, i_{2k}}$, $k >1$ may be written as a polynomial in $\wp_{ik}$ with coefficients expressible in terms of the invariants of the curve [BEL97]. A complete set of differential relations in the particular cases $g=2$ and $g=3$ can be found in \cite{ba03} \cite{ba07} and recently in covariant form these relations were obtained in \cite{athorne08}.
\er

We now restrict ourself to the case of a genus two curve, whose equation can be taken  in the form
\begin{eqnarray}
y^2&=&4x^5+\alpha_4x^4+\alpha_3x^3+\alpha_2x^2+\alpha_1x+\alpha_0\\
&=&4(x-a_1)(x-a_2)(x-a_3)(x-a_4)(x-a_5), \quad a_l\neq a_l.\label{genus2}
\end{eqnarray}

The holomorphic differentials $\mathbf{u}=(u_1,u_2)^T$ are related to
$ \vb=(v_1, v_2)^T$
\be {v}_1=\frac{x\mathrm{d} x}{y}, \quad  {v}_2=\frac{\mathrm{d} x}{y}  \ee
by the transformation
\begin{eqnarray}
T=\left(\begin{array}{cc}1&\frac{\alpha_4}{8}\\0&1\end{array}\right)
\end{eqnarray}

According to the definition (\ref{algbidifferentialexp}), the first few coefficients $\mu_{ij}^{\rm{alg}}$ are
\begin{eqnarray}
\mu_{ij}^{\rm{alg}}&=&0,\quad {\rm if}\; i\;\; {\rm or}\;\;j\;\; {\rm or}\;{\rm both}\;{\rm are}\; {\rm even},\cr
\mu_{1,1}^{\rm{alg}}&=&-\frac{1}{16}\alpha_4,\cr
\mu_{1,3}^{\rm{alg}}&=&\mu_{3,1}=-\frac{1}{16}\alpha_3+\frac{3}{256}\alpha_4^2,\label{mu}\cr
\mu_{1,5}^{\rm{alg}}&=&\mu_{5,1}=-\frac{1}{16}\alpha_2+\frac{3}{128}\alpha_3\alpha_4
-\frac{5}{2048}\alpha_4^3,\cr
\mu_{3,3}^{\rm{alg}}&=&-\frac{3}{16}\alpha_2+\frac{1}{32}\alpha_3\alpha_4
-\frac{3}{1024}\alpha_4^3,  \\
&\vdots& \nonumber
\end{eqnarray}

Introduce the set of residue vectors $\mathbf{R}_{2k}=0, k=1,\ldots$ and
\be
  \mathbf{R}_1=\left(\begin{array}{c}1\\0\end{array}\right),\;\; \mathbf{R}_3=\left(\begin{array}{c}-\frac{1}{8}\alpha_4\\1
\end{array}\right),\;\;
\mathbf{R}_5=\left(\begin{array}{c}-\frac{1}{8}\alpha_3 +\frac{3}{128}\alpha_4^2\\-\frac{1}{8}\alpha_4\end{array}\right), \ldots
\ee

It follows from formulae (\ref{schurderivhook}) and (\ref{schurderivbilin}) that the first few affine matrix components are

\begin{eqnarray}
A_{00}(\vb)&=&\zeta_1,\\
A_{01}(\vb)&=&\frac12\zeta_1^2-\frac12\wp_{11}-\frac{1}{16}\alpha_4\cr
A_{02}(\vb)&=&\frac16\zeta_1^3+\frac13\zeta_2-\left(\frac12\wp_{11}
+\frac{5}{48}\alpha_4 \right)\zeta_1-\frac16 \wp_{111},\cr
A_{03}(\vb)&=& \frac{1}{24}\zeta^4+\frac13\zeta_1\zeta_2
-\left(\frac{7}{96}\alpha_4+\frac14\wp_{11} \right)-\frac16\wp_{111}\zeta_1\cr
&-&\frac{1}{24}\wp_{1111}-\frac13\wp_{12}
+\frac18\wp_{11}^2+\frac{7}{96}\alpha_4\wp_{11}
-\frac{1}{24}\alpha_3+\frac{5}{512}\alpha_4^2,
\cr
&\vdots&\cr
&\vdots&\cr
A_{10}(\vb)&=&A_{01}(\vb),\cr
A_{11}(\vb)&=&\frac13\zeta_1^3-\frac13 \zeta_2-\left( \wp_{11}+\frac{1}{12}\alpha_4 \right)\zeta_1-\frac13 \wp_{111},\cr
A_{12}(\vb)&=&\frac18\zeta_1^4
-\frac12\zeta_1\wp_{111}+\frac38\wp_{11}^2-\frac18\wp_{1111}
-\left(\frac34\wp_{11}+\frac{3}{32}\alpha_4\right)\zeta_1^2+\frac{3}{32}\alpha_4\wp_{11}
+\frac{3}{512}\alpha_4^2,\cr
A_{13}(\vb)&=&\frac{1}{30}\zeta_1^5+\frac16\zeta_1^2\zeta_2
-\frac13\wp_{111}\zeta_1^2-\left(\frac13\wp_{11}+\frac{1}{16}\alpha_4 \right)\zeta_1^3-\left( \frac16\wp_{11}+\frac{1}{48}\alpha_4 \right)\zeta_2\cr
&+&\left( -\frac16\wp_{1111}-\frac13\wp_{12}+\frac12\wp_{11}^2
+\frac{3}{16}\alpha_4\wp_{11}-\frac{1}{24}\alpha_3+\frac{7}{384}\alpha_4^2 \right)\zeta_1\cr
&+& \frac13\wp_{11}\wp_{111}-\frac16\wp_{1112}-\frac{1}{30}\wp_{11111}
+\frac{1}{16}\alpha_4\wp_{111},\cr
&\vdots&\nonumber
\end{eqnarray}
All Kleinian symbols, $\zeta_{k}, \wp_{k,l}$ etc. in these formulae are evaluated at $\vb$.

\bp
\label{3_2_identities_prop}
The Pl\"ucker relations written for the partition $\lambda=(2,2)$
of  weight 4  and partitions $\lambda=(3,2),\lambda= (2,2,1)$ of  weight 5 are equivalent to the equation
\begin{eqnarray} \wp_{1111}(\vb)=6\wp_{11}^2(\vb)+4\wp_{12}(\vb)
+\alpha_4\wp_{11}(\vb)+\frac12\alpha_3.\label{kdv1}\end{eqnarray}
\ep

\noindent{\bf Proof}. The first non-trivial Pl\"ucker relation for the partition $\lambda=(2,2)$
\begin{eqnarray}
\pi_{(1,0|1,0)}=\left|\begin{array}{cc} A_{1,1}&A_{1,0}\\
A_{0,1}&A_{0,0} \end{array} \right| \label{giambelli2}
\end{eqnarray}
is written in detailed form as
\begin{eqnarray}
&&\tau(\mathbf{0},\vb)\left.\left[
\frac{1}{12}\frac{\partial^4}{\partial t_1^4}+\frac14 \frac{\partial^2}{\partial t_2^2}-\frac13 \frac{\partial^2}{\partial t_1\partial t_3}
\right] \tau(\mathbf{t},\vb)\right|_{\mathbf{t}=0}\cr\cr\cr&&\left|\begin{array}{cc}
\left.\displaystyle{\left[\frac13\frac{\partial^3}{\partial t_1^3}-\frac13\frac{\partial}{\partial t_3}\right]
\tau(\mathbf{t},\vb)}\right|_{\mathbf{t}=0}
&\left.\displaystyle{\left[\frac12 \frac{\partial}{\partial t_2}+\frac{\partial^2}{\partial t_1^2} \right]\tau(\mathbf{t},\vb)}\right|_{\mathbf{t}=0}\\\\
\left.\displaystyle{\left[-\frac12\frac{\partial}{\partial t_2} +\frac12
\frac{\partial^2}{\partial t_1^2}\right]} \tau(\mathbf{t},\vb)\right|_{\mathbf{t}=0}&\left.\displaystyle{\frac{\partial}{\partial t_1}\tau(\mathbf{t},\vb)}\right|_{\mathbf{t}=0}
\end{array}\right|
 \end{eqnarray}

Substituting the expression (\ref{tau_sigmae}) for $\tau(\mathbf{t};\eb)$ into this relation
and expressions for $\mu_{ij}$ and computing directional derivatives, we obtain equation (\ref{kdv1}). Expressions for the hook diagram coefficients entering in the determinant are given above and
\bea
\pi_{(1,0|1,0)}&=&-\frac{1}{12}\wp_{1111}+\frac{1}{12}\zeta_1^4
-\left(\frac12\wp_{11}+\frac{1}{48}\alpha_4\right)\zeta_1^2-\frac13\zeta_2\zeta_1 -\frac13\wp_{1 11}\zeta_1\label{pi1010}\\
&+&\frac13\wp_{12}+\frac14\wp_{11}^2
+\frac{1}{48}\alpha_4\wp_{11}-\frac{1}{256}\alpha_4^2+\frac{1}{24}\alpha_3.\nonumber
\eea
Relation (\ref{kdv1}) follows from these.

 The partitions $\lambda=(3,2)$ and $\lambda=(2,2,1)$ of weight 5  give
Pl\"ucker relations that imply the action of $\zeta_1(\vb)+ \partial/\partial v_1$ on the above equation. These all yield the single equation.
\[ \left\{\begin{array}{c} \yng(2,2)\\
 \yng(3,2) \quad \yng(2,2,1)\end{array}\right. \Leftrightarrow  \wp_{1111}(\vb)=6\wp_{11}^2(\vb)+4\wp_{12}(\vb)
+\alpha_4\wp_{11}(\vb)+\frac12\alpha_3 .\]

Analogous considerations involving weight 6 and 7 partitions lead to the correspondence
\begin{eqnarray*} &&\left\{\begin{array}{c} \yng(4,2)\; \quad
\yng(3,3)\quad \quad \yng(3,2,1)\quad\yng(2,2,2)\quad\yng(2,2,1,1)\\
\yng(5,2)\quad \ldots\ldots\ldots\ldots\ldots\ldots\ldots\ldots \quad \yng(2,2,1,1,1)
\end{array}\right.
\end{eqnarray*}
\begin{eqnarray*} \Updownarrow  \end{eqnarray*}
\begin{eqnarray}
\wp_{1112}(\vb)&=&6\wp_{11}\wp_{12}(\vb)-2\wp_{22}(\vb)
+\alpha_4\wp_{12}(\vb)\label{kdv2}\\
\wp_{111}^2(\vb)&=&4\wp_{11}^3(\vb)+\wp_{22}(\vb)
+4\wp_{12}(\vb)\wp_{11}(\vb)+\alpha_4\wp_{11}^2(\vb)
+\alpha_3\wp_{11}(\vb)\label{jac6}
\end{eqnarray}
Using the definition of weight $\mathcal{W}$ for the functions appearing in
(\ref{kdv2}) and (\ref{jac6}) we see that all these equations are homogeneous.

The process described here can be continued.  It seems reasonable to conjecture that the whole set of differential relations between multi-index Kleinian symbols can be put into correspondence with the Young diagrams of the partitions $\lambda=(2,2,i_1,\ldots,i_n)$ in such the way that partitions of  weights $2k$ and $2k+1$ correspond to a set of equations of weight $\mathcal{W}=2k$.

To complete the interpretation of the basic equations describing Abelian functions in terms of Pl\"ucker relations, we find that the Kummer surface arises as the Pl\"ucker relation corresponding to the
$\lambda=(4,4,4,4)$ diagram with weight 16.
\be
\pi_{(3210|3210)}=\left| \begin{array}{cccc}
A_{3,3}&A_{3,2}&A_{3,1}&A_{3,0}\\
A_{2,3}&A_{2,2}&A_{2,1}&A_{2,0}\\
A_{1,3}&A_{1,2}&A_{1,1}&A_{1,0}\\
A_{0,3}&A_{0,2}&A_{0,1}&A_{0,0}  \end{array}\right|. \label{kummer}
\ee
The equation
\be \pi_{(3210|3210)}=0 \label{kummer1} \ee
can be written in the form
\be \left |  \begin{array}{cccc}  \alpha_0&  \frac{1}{2}
 \alpha_1&-2  \wp_{22}&-2  \wp_{12}  \\
  \frac{1}{2}  \alpha_1&  \alpha_2+4  \wp_{22}&  \frac{1}{2}
\alpha_3+ 2  \wp_{12}&-2  \wp_{11}  \\-2  \wp_{22}&  \frac{1}{2}
 \alpha_3+2  \wp_{12}&  \alpha_4+4  \wp_{11}&2  \\
-2  \wp_{12}&-2  \wp_{11}&2&0  \end{array}  \right|=0.\label{kummer3}\ee
This is the celebrated quartic Kummer surface, $\mathrm{Kum} (X)$, defined as  the surface in $\mathbb{C}^3$ with coordinates $x=\wp_{11},y=\wp_{12},z=\wp_{22}$.
 $\mathrm{Kum} (X)$ is the quotient of the Jacobi variety,
 $\mathrm{Kum}(X)=\mathrm{Jac}(X)/(\ub\rightarrow - \ub )$

Therefore we conclude
\begin{eqnarray*}
\yng(4,4,4,4)\qquad \begin{array}{c}\Longleftrightarrow \\ \phantom{a}\\
\phantom{a} \\ \phantom{a} \end{array}\quad \begin{array}{c}\mathrm{Kum} (X)\\
\phantom{a}\\
\phantom{a}\\ \phantom{a}  \end{array}
\end{eqnarray*}

\br
The equation for the Kummer surface in this form was derived by Baker \cite{ba07} and a generalization to higher genera was given in \cite{bel97b}.
Also note that equations written for all $2\times 2$ minors of the matrix
$(A_{i,j})_{i,j=0,\ldots,3}$ in (\ref{kummer}),
\be  \pi_{(i,k|j,l)} =\left|\begin{array}{cc} A_{i,j}&A_{i,l}\\
                                              A_{k,j}&A_{k,l} \end{array}  \right| ,\quad i\geq j,\quad k\geq l, \quad i,j,k,l\in\{0,1,2,3\} \ee
\er
give a complete set of algebraic equations describing the Jacobi variety  $\mathrm{Jac}(X)$ as an
algebraic variety and also the flows of KdV type on $\mathrm{Jac}(X)$. This resembles the matrix realization of the Jacobi and Kummer varieties given by Baker \cite{ba07}, which was generalized to higher genera in \cite{bel97b}.

The above considerations lead to the following result.

\bt Each column vector $\mathbf{A}_k(\boldsymbol{\vb})$, $k=0,\ldots$ in the matrix of affine
coordinates of the element of the Grassmanian, whose components  $A_{k,l}$ $l=1,\ldots$ correspond to hook  partitions $(k+1,1^l) $ is a polynomial in a finite set of Kleinian symbols $\zeta_i(\vb),\ \wp_{ij}(\vb),\ \wp_{ijk}(\vb)$.
\et


\subsection{$\tau$-function of a trigonal curve}
In this section we demonstrate how the above results appear
in the case of a trigonal curve. The $\sigma$-function theory of trigonal Abelian functions was developed in \cite{bel00}. Various results in this area were obtained in \cite{eel00}, \cite{bl05}, \cite{eemop07},
\cite{matprev08}, \cite{ee09}, \cite{matprev10}.
In order to emphasize the main idea  and avoid  cumbersome formulae we restrict ourselves to the first nontrivial case of the cyclic family of trigonal curves $X$ of genus $3$ defined by:
\be
P(x,y)=y^3 -(x^4 +\beta_3 x^3 +\beta_6 x^2 +\beta_9 x +\beta_{12})=0.\label{trig}
\ee
and fix a canonical basis of cycles
$(\mathfrak{a}_1,\ldots,\mathfrak{a}_3;\mathfrak{b}_1,
\ldots,\mathfrak{b}_3)\in H_1(X,\mathbb{Z})$ of $X$.

The explicit calculation of canonical holomorphic differentials and
the meromorphic differentials conjugate to them is given in \cite{eemop07}.
In particular we have for $p=(x,y)$
\begin{eqnarray*}
&&u_1(p)=\frac{\mathrm{d}x}{3y},\quad
u_2(p)=\frac{x\mathrm{d}x}{3y^2},\quad
u_3(p)=\frac{\mathrm{d}x}{3y^2},\quad
\\
\\
&&r_1(p)=\frac{x^2\mathrm{d}x}{3y^2},\qquad
r_2(p)=-\frac{2x\mathrm{d}x}{3y},\qquad
r_3(p)=-\frac{(5x^2+3\beta_3x+\beta_6)\mathrm{d}x}{3y}.
\end{eqnarray*}
Denote, as above, the period matrices, $\grA$, $\grB$ and $\boldsymbol{T}=\grA^{-1}\grB$, $\varkappa=\grA^{-1}\grS$.

Let $\mathcal{D}=p_1+p_2+p_3$ be a nonspecial divisor of degree $3$ and
\be
 \mathbf{v}=\sum_{i=1}^3\int_{p_{\infty}}^{p_i} \mathbf{u}+\grA\mathbf{K},\label{Abeltrig}
 \ee
where $\mathbf{K}$ is the vector of Riemann constants with the base point at $p_{\infty}$.

The polynomial $\mathcal{F}((x,y);(z,w))$ appearing in the fundamental bi-differential $\Omega(p,q)$  is given by
\begin{eqnarray}
\mathcal{F}\big((x,y),(z,w)\big)=3w^2 y^2+wT(x,z)+yT(z,x),
 \label{omegatrig1}
\end{eqnarray}
where
\begin{eqnarray}
    T(x,z)=3\beta_{12}+ ( z+2x ) \beta_9+x ( x+2\,z )\beta_6
    +3\beta_3 x^2 z + x^2 z^{2}+2\,x^3z.\label{omegatrig2}
\end{eqnarray}

Expanding about $p_{\infty}$ gives the following expressions for the quantities $\mu_{ij}^{\rm{alg}}$
\begin{eqnarray}
\mu_{0,0}^{\rm{alg}}&\&=0,\\
\mu_{0,1}^{\rm{alg}} &\&=\mu_{1,0}^{\rm{alg}}=-\frac{2}{3}\beta_3,\cr
\mu_{0,4}^{\rm{alg}}&\& =\mu_{4,0}^{\rm{alg}}=-\frac{2}{3}\beta_6+\frac{5}{9}\beta_3^2,\cr
\mu_{1,3}^{\rm{alg}}&\&=\mu_{3,1}^{\rm{alg}}=-\frac{2}{3}\beta_6+\frac{4}{9}\beta_3^2,\cr
\mu_{2,2}^{\rm{alg}}&\&=0, \cr
\vdots \nonumber
\end{eqnarray}
\begin{remark} $\mu_{ij}^{\rm{alg}}=0$ unless $(i+j)+2\equiv 0\ \mathrm{mod}\, 3$. This is a consequence of the cyclic symmetry of the curve. \end{remark}

In this case the Klein formula  reads
\begin{eqnarray*}
\sum_{i,k=1}^3\wp\left( \int_{p_{\infty}}^p \mathbf{u} -\vb\right)\phi_i(x,y)\phi_k(x_r,y_r)
=\frac{\mathcal{F}(p,p_r)}{(x-x_r)^2},\quad r=1,2,3,
\end{eqnarray*}
where $p=(x,y)$, $p_k=(x_k,y_k)$ also $\phi_1(x,y)=y, \phi_2(x,y)=x,\phi_3(x,y)=1$.

Expanding this relation in the vicinity of  $p_{\infty}$, where local coordinate $x=1/\xi^3$ is introduced/. Equating principal parts at the poles, we obtain a set of equations involving variables $x_x,y_k$ and $\wp$-symbols. We now show that these relations can be obtained as consequences of
Pl\"ucker relations via Giambelli-type formula.

The first Young diagrams leading to a nontrivial Pl\"ucker relation, as in the hyperelliptic genus 2 case, corresponds to the partition $\lambda=(2,2)$.
In this case we obtain, after simplification, the equation
\be
\wp_{1111}= 6\, \wp_{11}^{2}-3\,\wp_{22}. \label{bous}
\ee
Differentiating with respect to the coordinate $v_1$ gives the Boussinesq equation.

The derivation of (\ref{bous}) is based on formula (\ref{giambelli2})  for the trigonal curve (\ref{trig}). Namely, we have
\be
\pi_{(1,0|1,0)}=\frac14\wp_{22}+\frac14\zeta_2^2-\frac{1}{12}\wp_{1111}-\frac13\wp_{11 1}\zeta_1+\frac14\wp_{11}^2-\frac12\wp_{11}\zeta_1^2+\frac{1}{12}\zeta_1^4,
\label{pi1010t}
\ee
and also
\bea
A_{0,0}(\vb)&=&\zeta_1(\vb),\nonumber\\
A_{0,1}(\vb)&=&-\frac12\wp_{11}(\vb)+\frac12\zeta_1^2(\vb)-\frac12\zeta_2(\vb),\label{Atrig}\\
A_{1,0}(\vb)&=&\frac12\zeta_2(\vb)-\frac12\wp_{11}(\vb)+\frac12\zeta_1^2(\vb),\nonumber\\
A_{1,1}(\vb)&=&-\frac13\wp_{111}(\vb)-\wp_{11}(\vb)\zeta_1(\vb)+\frac13\zeta_1^3(\vb).\nonumber
\eea
In the trigonal case we no longer have symmetry about the
diagonal of the Young diagram that we have in the genus 2 case, but we can
restrict ourselves to equations of even degree by taking  symmetric
combinations of the two diagram related by transposition.  In the
weight 5 case we have the symmetric combination
\[
\yng(3,2)\quad + \quad \yng(2,2,1),
\]
which gives the weight 5 trigonal PDE
\[
\wp_{1112}=6\,\wp_{11}\wp_{12}+3\,\beta_{3}\wp_{11}.
\]
For weight 6 we have the three sets of diagram
\begin{eqnarray*}
  \yng(4,2)\quad + \quad\yng(2,2,1,1), \quad \yng(3,3)\quad
  +\quad\yng(2,2,2)\,, \quad \yng(3,2,1),
\end{eqnarray*}
which lead to an over-determined set of equations with the unique
solution
\begin{eqnarray*}
  \wp_{111}^{2} & =4 \wp_{11} ^{3}+
  \wp_{12}^{2}+4\,\wp_{13}-4\,\wp_{11}\wp_{22},\\
  \wp_{1122}& =4\,\wp_{13}+4\wp_{12}^{2}+
  2\,\wp_{11}\wp_{22}+3\,\beta_{3}\wp_{12}+2\,\beta_{6}.
\end{eqnarray*}
Continuing in this way, we recover the cyclic trigonal versions of
the full set of equations given in \cite{eemop07}.

{\bf Acknowledgments} This work was partially supported by the European
Science Foundation Programme MISGAM (Methods of Integrable System,
Geometry and Applied Mathematics).  The authors are grateful
to the HWK, Hanse-Wissenschaftskolleg (Institute of Advanced Study) in Delmenhorst
for a grant enabling them to work together in May 2010 at HWK to complete the final version of this paper. Work of J.H. was supported in part by the Natural Science and Engineering Research Council at Canada (NSERC) and the Fonds Qu\'ebe\'cois de la recherche sur la nature et les technologies (FQRNT). The authors are also grateful to Luc Haine for helpful discussions and for bringing to our attention the
unpublished manuscript by J.\ Fay \cite{fay83}.

\providecommand{\bysame}{\leavevmode\hbox to3em{\hrulefill}\thinspace}
\providecommand{\MR}{\relax\ifhmode\unskip\space\fi MR }
\providecommand{\MRhref}[2]{%
  \href{http://www.ams.org/mathscinet-getitem?mr=#1}{#2}
}
\providecommand{\href}[2]{#2}

\end{document}